\PassOptionsToPackage{combine}{ucs}
\documentclass[pre, aps,twocolumn,superscriptaddress,longbibliography,floatfix,notitlepage]{revtex4-2}

\usepackage[utf8]{inputenc}
\usepackage[english]{babel}
\usepackage[T1]{fontenc}
\usepackage{lmodern}
\usepackage{amsmath,amsfonts,amssymb,amsthm,bm,times}
\usepackage[colorlinks={true}, citecolor={blue}, filecolor={blue}, linkcolor={blue}, urlcolor={blue}]{hyperref}
\usepackage{graphicx,color}
\usepackage[caption=false]{subfig}
\usepackage{microtype}
\usepackage{braket}
\usepackage{gensymb} 
\usepackage{float}
\DeclareUnicodeCharacter{2212}{-}

\begin{document}

\title{Few-qubit quantum refrigerator for cooling a multi-qubit system}

\author{Onat Ar\i soy}
\affiliation{Institute for Physical Science and Technology, University of Maryland, College Park, Maryland 20742, USA}
\affiliation{Chemical Physics Program, University of Maryland, College Park, Maryland 20742, USA}
\author{\"{O}zg\"{u}r E.\ M\"{u}stecapl{\i}o\u{g}lu}
\email[Electronic address:\ ]{omustecap@ku.edu.tr}
\affiliation{Department of Physics, Ko\c{c} University, Sar{\i}yer, \.Istanbul, 34450, Turkey}
%
%
%

\date{\today}

\begin{abstract}
We propose to use a few-qubit system as a compact quantum refrigerator for cooling an interacting multi-qubit system. We specifically consider a central qubit coupled to $N$ ancilla qubits in a so-called spin-star model as our quantum refrigerator. We first show that if the interaction between the qubits is of the longitudinal and ferromagnetic Ising model form, the central qubit is colder than the environment.  The colder central qubit is then proposed to be used as the refrigerant interface of the quantum refrigerator to cool down general quantum many-qubit systems. We discuss a simple refrigeration cycle, considering the operation cost and cooling efficiency, which can be controlled by $N$ and the qubit-qubit interaction strength. Besides, bounds on the achievable temperature are established. Such few-qubit compact quantum refrigerators can be significant to reduce dimensions of quantum technology applications, can be easy to integrate into all-qubit systems, and can increase the speed and power of quantum computing and thermal devices. 
\end{abstract}

\maketitle


\section{\label{sec:Intro}Introduction}
The widespread use of quantum technologies is limited by the large and expensive cooling systems required for their implementations. The rapidly emerging field of quantum thermodynamics~\cite{landi_irreversible_2020,kosloff_quantum_2013,partovi_quantum_1989,ozdemir_quantum_2020,vinjanampathy_quantum_2016} paves the way for compact, fast, and efficient quantum refrigeration schemes for quantum devices~\cite{naseem_two-body_2020,abah_shortcut-adiabaticity_2020}. Pioneering studies are limited to cooling a single two-level system 
(qubit or spin-$1/2$ particle)~\cite{pulse}. A critical question for practical quantum machines is if and how such quantum refrigerators can cool down interacting multiple qubit systems. As a possible positive answer to this question, we propose a few-qubit quantum refrigerator with scalable advantages in its cooling efficiency and achievable minimum temperatures. 

Early quantum refrigerator studies consider utilization of quantum coherence injected by external drives~\cite{pulse}, spectral bath filtering and periodically modulated interactions~\cite{drivenref}, and frequent measurement schemes~\cite{meas}.
The requirements of such proposals, precise quantum control~\cite{pulse,drivenref}, bath engineering~\cite{drivenref,meas}, very rapid measurements~\cite{meas} together with the lack of precise determination of energetic costs of quantum control and measurements make them difficult to implement for practical applications. A more conventional cooling method for spin systems is known as algorithmic cooling~\cite{alg1,alg2,alg3}. How it can be part of a quantum algorithmic heat engine has been recently presented~\cite{kose_algorithmic_2019}. A continuous variant of algorithmic cooling, without an external work cost, allowing for a flexible working temperature range, is proposed~\cite{smallref}. However, it relies on a three-body interaction among the qubits, which is not feasible for experimental realization. Intriguing proposals based on quantum coherence and entanglement to cool quantum systems~\cite{lutz2009,coh3} are challenging to use in refrigeration cycles. Their cost to prepare entangled states repeatedly reduces their appeal for practical applications.

Recently, a scheme, closely related to algorithmic cooling idea of entropy transfer between different qubit systems, to thermalize a many-body system by repeated collisions has been proposed~\cite{ourpaper}. The random collisions are
one of the oldest routes considered for describing thermalization, introduced by Lord Rayleigh~\cite{rayleigh1891}. A massive
particle thermalizes after many random collisions by small projectiles in
thermal states. This mechanism explains the micromaser in
the blackbody radiator regime, where
the optical cavity is heated by thermal pump atoms~\cite{scully_quantum_1967}. 
More recent studies showed that pump atoms in quantum coherent states could also be used to heat the micromaser~\cite{coh1,coh2,epl-pce}. A particularly
intriguing scenario is the scalable heating of the micromaser with the number of pump atoms, using a so-called spin-star system~\cite{epl-pce}. Spin-star configuration consists of a central qubit surrounded by $N$ ancilla qubits (cf.~Fig.~\ref{fig:spinstar}). The critical point is that the central qubit can be at a higher local temperature than the environment.

Here, we show that when the interaction between the central spin and the surrounding spins in a central spin model contains only the longitudinal spin components and is of ferromagnetic type (negative coupling coefficient), the central spin becomes locally colder than the environment. Accordingly, the $(N+1)$-qubit system can be envisioned as a quantum refrigerator, where the central qubit is the quantum refrigerant to cool other systems, specifically, an interacting
multi-qubit system. For that aim, it is necessary to contact the quantum refrigerant with the many-body system. The required refrigerant-system coupling can be performed within the collisional route to many-body thermalization~\cite{ourpaper}. Successful coupling needs matching refrigerant frequencies to transition frequencies of the many-body system. Therefore, our proposal can be envisioned as an all-qubit network with integrated quantum refrigerators (cf.~Fig.~\ref{fig:coll}). 

Finally, we should clarify the similarities with the algorithmic cooling. There is only a single bath (environment) where the spin-star qubit structure is held. Such quantum "molecule" has a central qubit at a local thermal equilibrium colder than the environment, due to longitudinal ferromagnetic qubit-qubit "bonding". While the initial thermal states' preparation is relatively easy in our scheme, we still need resetting and timing control in the quantum cooling network. Similar to algorithmic cooling, timing and control can be achieved by using qubits at different thermalization rates. Another significant advantage here is to have a readily integrable few-qubit refrigerator with a single qubit refrigerant for compact, fast, and efficient cooling of a many-qubit system.  

While our focus will be on thermalization with the Markovian collision model introduced in Ref.~\cite{ourpaper} for the rest of this paper, another recent work on Markovian collision models for many-body systems \cite{mark-coll} needs to be mentioned. Although it is based on couplings much stronger than the system Hamiltonian, constraining its range of possible implementations, it is promising to generate non-local Lindblad dissipators, which are necessary to thermalize many-body systems with non-local energy eigenstates, using multi-qubit quantum gates. The collision model of Ref.~\cite{ourpaper} is constrained to local collisions and it is not guaranteed to generate a Lindblad master equation with a positive definite Kossakowski matrix for many-body systems with entangled energy eigenstates.

The outline of our paper is the following. After giving a brief description of our spin-star model in Sec.~\ref{mod}, we will derive an analytical expression for the effective temperature of the central qubit with uniform Ising interaction between center and ancilla qubits in Sec.~\ref{th}. After working out how cold these interactions can get the central qubit, we will discuss a simple refrigeration cycle in Sec.~\ref{sec:refrigeratorCycle} and calculate its efficiency defined as the ratio of the energy taken from the central qubit to the total work spent in one cycle. The Section~\ref{many} will summarize the findings of our previous work on many-body systems~\cite{ourpaper} and clarify how it allows cooling of quantum many-body systems along with this paper. The Section \ref{ancilla-sec} will deal with the state of the ancilla qubits after thermalization and we will propose two possible ways to use the ancilla qubits to make our refrigerator proposal more efficient. We conclude in Sec.~\ref{sec:conclusion}. We investigate the quantum effects in our refrigerator with Heisenberg interaction and provide a brief quantum-classical comparison for our model in the appendix. 

\section{Model system\label{mod}}

\begin{figure}[t!]
  \centering
\includegraphics[width=\linewidth]{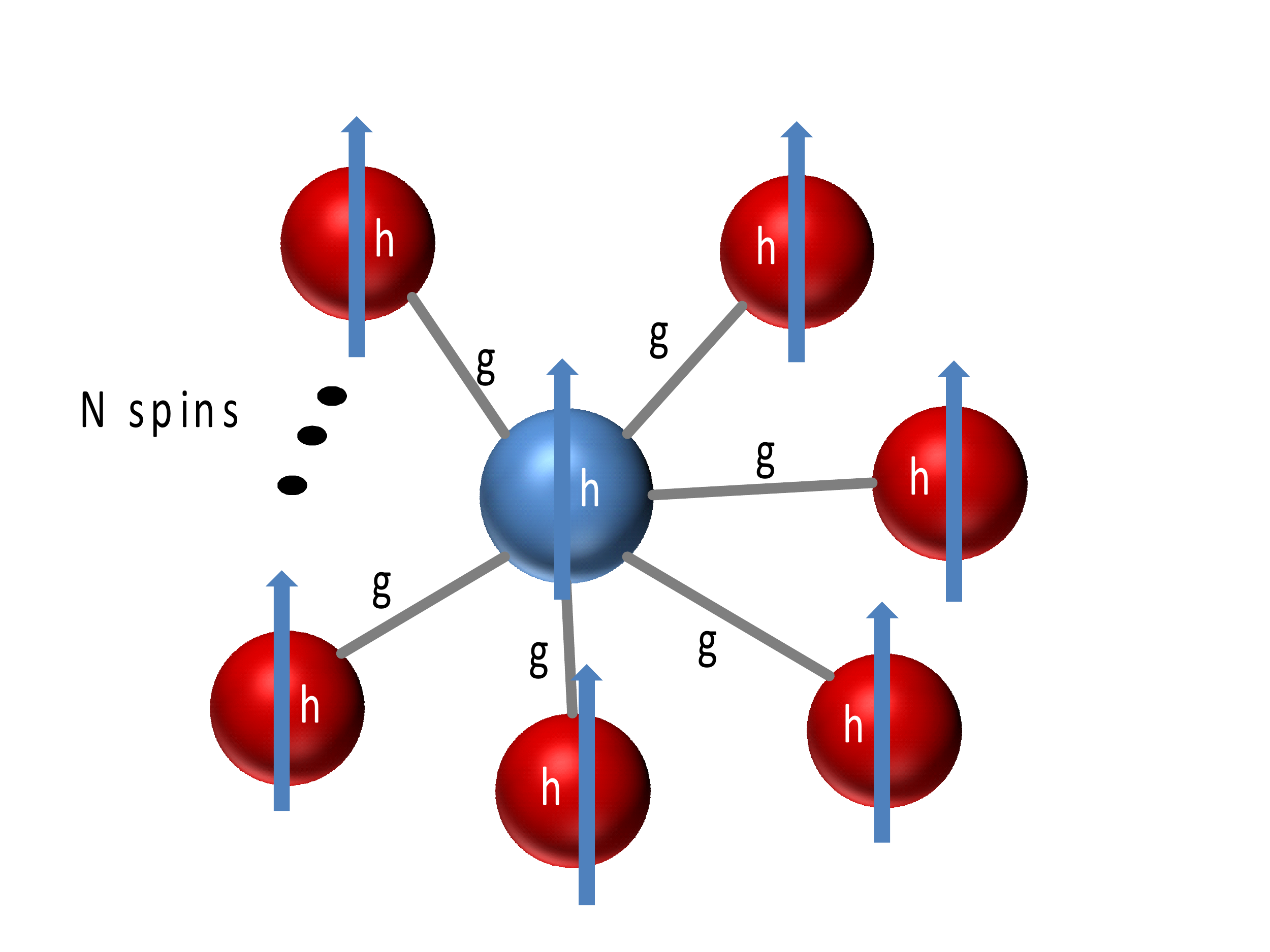}
\caption{Sketch of spin-star model consisting of a central spin-$1/2$ particle (blue sphere) surrounded by $N$ spin-$1/2$ particles (red spheres). Central spin is coupled to the surrounding spins with the same interaction coefficient $g$. The whole system is
	in a homogeneous magnetic field $h$. We assume the interactions only contains longitudinal spin components, in the same direction with the magnetic field.  The spin-star model is used to describe a $(N+1)$-qubit quantum refrigerator, where each spin effectively represents a qubit with an energy gap $h$. When the system is in thermal equilibrium with an environment at temperature $T$, the central qubit is at a temperature smaller than $T$. The central qubit can be used as the refrigerant to cool other quantum many-body systems (cf.~Fig.~\ref{fig:coll}).}
\label{fig:spinstar}
\end{figure}

We consider a so-called "spin-star" system consisting of a single qubit surrounded by $N$ ancilla qubits, as illustrated in Fig.~\ref{fig:spinstar}. Interactions between the central qubit and the surrounding qubits are assumed to be the same, characterized by the coupling coefficient $g$. The energy gap of the qubit is denoted by $h$. We represent each qubit as an effective spin-$1/2$ particle and further assume that the qubit-qubit interactions are only between the $z$-components of the effective spins. The specification of interaction direction is neither for simplicity nor arbitrary. Transverse components cause correlations and entanglement in the eigenstates of the Hamiltonian, which is not desirable for our purpose of cooling the system. Further explanation of the harmful influence of transverse interactions on cooling is given in the appendix. The total Hamiltonian can be written as
\begin{equation}\label{eq:model}
\hat{H} = h\sum_{n=0}^{N}\hat{\sigma}_{z,n} + g~\hat{\sigma}_{z,0}\sum_{n=1}^{N}\hat{\sigma}_{z,n},
\end{equation}
where the indices $n=0$ and $n=1,2\dots N$ indicate the central qubit  and surrounding qubits, respectively. $\hat\sigma_{z,0},\hat\sigma_{z,n}$ are the $z$-component Pauli spin operators.

As the Pauli spin operators are only for the $z$-components, the model can be considered a longitudinal Ising model~\cite{ising}, but with a spin-star configuration instead of a spin chain. Spin-star models are special cases of Richardson-Gaudin models, which are usually studied in the context of hyperfine interactions in semiconductor quantum dots~\cite{qdot1,qdot2} and as a toy model of non-Markovianity~\cite{nm-ss1,nm-ss2,nm-ss3}. However, the semiconductor quantum dot implementation of spin-star models will not be relevant for our purposes. It is based on Heisenberg interactions, which we discuss and rule out for our purposes in the appendix. For a superconducting qubit implementation of our proposal, a generalization and re-configuration of the Chimera unit cell architecture used in D-Wave quantum annealers seems possible. This architecture makes use of orthogonally placed qubits overlapping each other and allowing to set couplers between horizontal and vertical qubits, which generate a longitudinal Ising interaction as we desire~\cite{dwave}. 
 
\section{Results}
\label{sec:results}
In our numerical simulations, we will consider an artificial spin system, specifically a system of superconducting two-level systems (qubits). Efficient, compact, and fast cooling of such superconducting interacting qubits is a critical problem for practical quantum computations. Hence, we focus our range of parameters on this particular case, though our generic models, exact analytical results, and general conclusions apply a broader class of physical systems. We will call "effective spin," representing a qubit as "spin" in the following discussions for brevity. We take $\hbar=1$ and set $h=1\text{ GHz}$ for all of our calculations as it is a typical order of magnitude for superconducting qubits~\cite{s-qubit} and we will assume that $g$ can be in the order of $h$~\cite{coupling}.
\subsection{Thermal state for the spin-star model and effective temperature of the center qubit\label{th}}
\label{sec:results-centerSpinTemp}
The eigenstates of the Hamiltonian in Eq.~(\ref{eq:model}) are not entangled. Off-diagonal elements of the total density matrix vanish in the tensor product of the z-basis of each effective spin. Accordingly, we can treat the Ising spin-star model as a classical discrete system (with up and down spin states labeled by $z=+1$ and $z=+1$, respectively) and study the state probability distribution described by the diagonal elements of the total density matrix.

We consider the spin-star system immersed in a thermal environment at inverse temperature $\beta$ which is related to the environment temperature by $\beta = 1/k_B T$. We can define the partition function of the whole system by treating up and down states of the central spin separately. Assuming the central spin is in the $z_0=\pm 1$ state, the partition function of a single ancilla spin equals to that of a non-interacting spin with Hamiltonian eigenvalue $h\pm g$.  The partition function of all ancilla spins is obtained simply by taking $N^{\text{th}}$ power of the partition function of a single ancilla. Summing the partition functions of ancilla spins for up (down) states of the central spin with factors $\exp{(-(+)\beta h)}$, we find the partition function of the whole system to be

\begin{equation}
Z_{\text{tot}} = 2^{N}(e^{-\beta h}\cosh^N(\beta(g+h))+e^{\beta h}\cosh^N(\beta(h-g))).\label{ztot}
\end{equation}
The first term of Eq. (\ref{ztot}) corresponds to the up state of the central spin while the second corresponds to its down state. That remark allows us to give explicit expressions for the probabilities of the states of the central spin
\begin{eqnarray}\label{eq:populations}
P(z_0 = \pm 1) = \frac{2^N e^{\mp \beta h} \cosh^N(\beta(h\pm g))}{Z_{\text{tot}}}.
\end{eqnarray}
The effective (local) inverse temperature of the central qubit $\beta_\text{eff}$ as a function of its state populations is defined by
\begin{eqnarray}\label{betaeff}
\beta_{\text{eff}} &=& \frac{1}{2h}\ln\left(\frac{P(z_0 =-1)}{P(z_0 =1)}\right) \nonumber\\
&=&\frac{1}{2h}\left(2\beta h + N\ln\left(\frac{\cosh(\beta(h-g))}{\cosh(\beta(h+g))}\right)\right)  \nonumber\\
&=& \beta + \frac{N}{2h}(\ln(\cosh(\beta(h-g)))-\ln(\cosh(\beta(h+g)))). \nonumber \\
\end{eqnarray}

\begin{figure}[t!]
	\centering
	\subfloat[\label{fig:2a}]{\includegraphics[width=\linewidth]{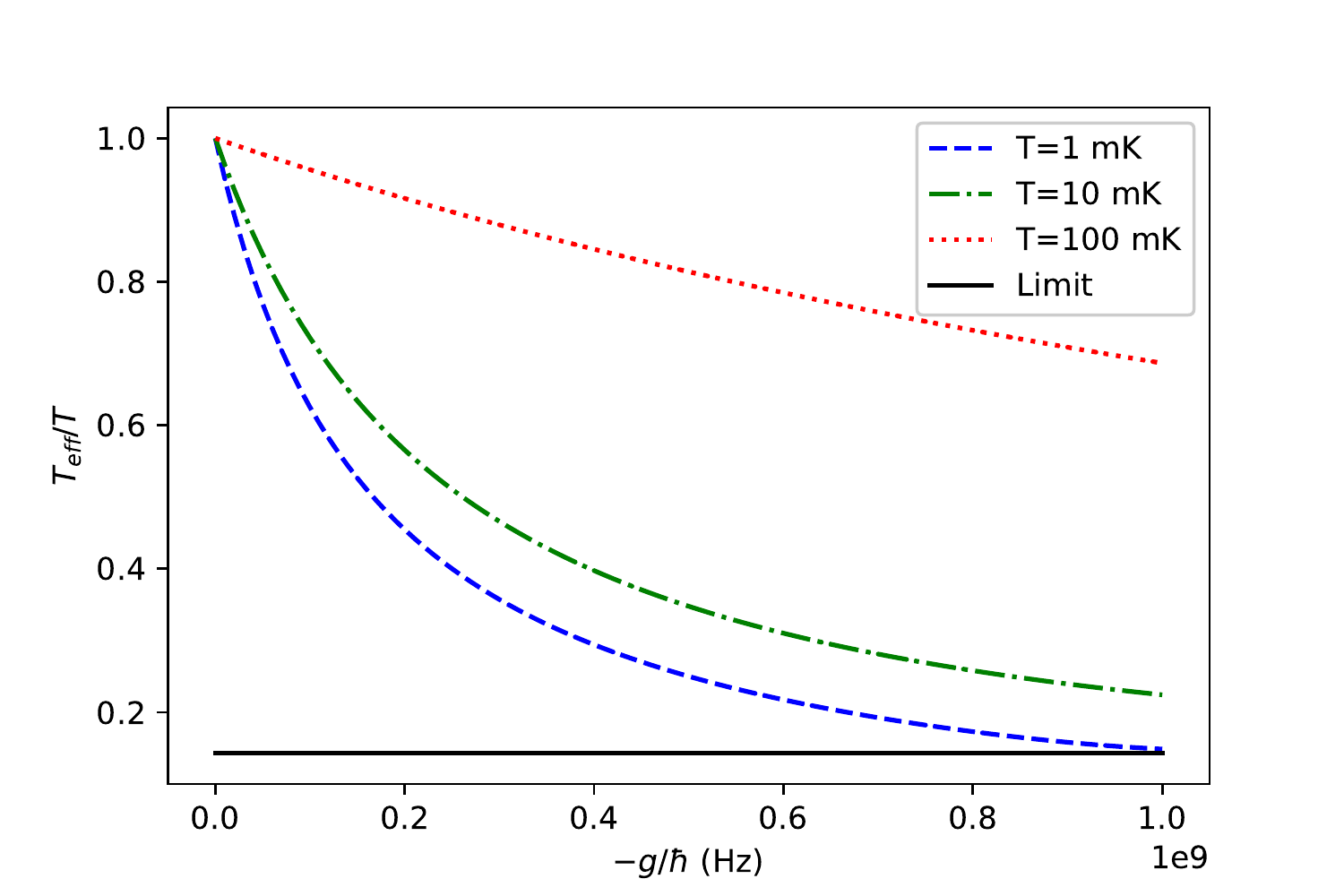}}
	\qquad
	\subfloat[\label{fig:2b}]{\includegraphics[width=\linewidth]{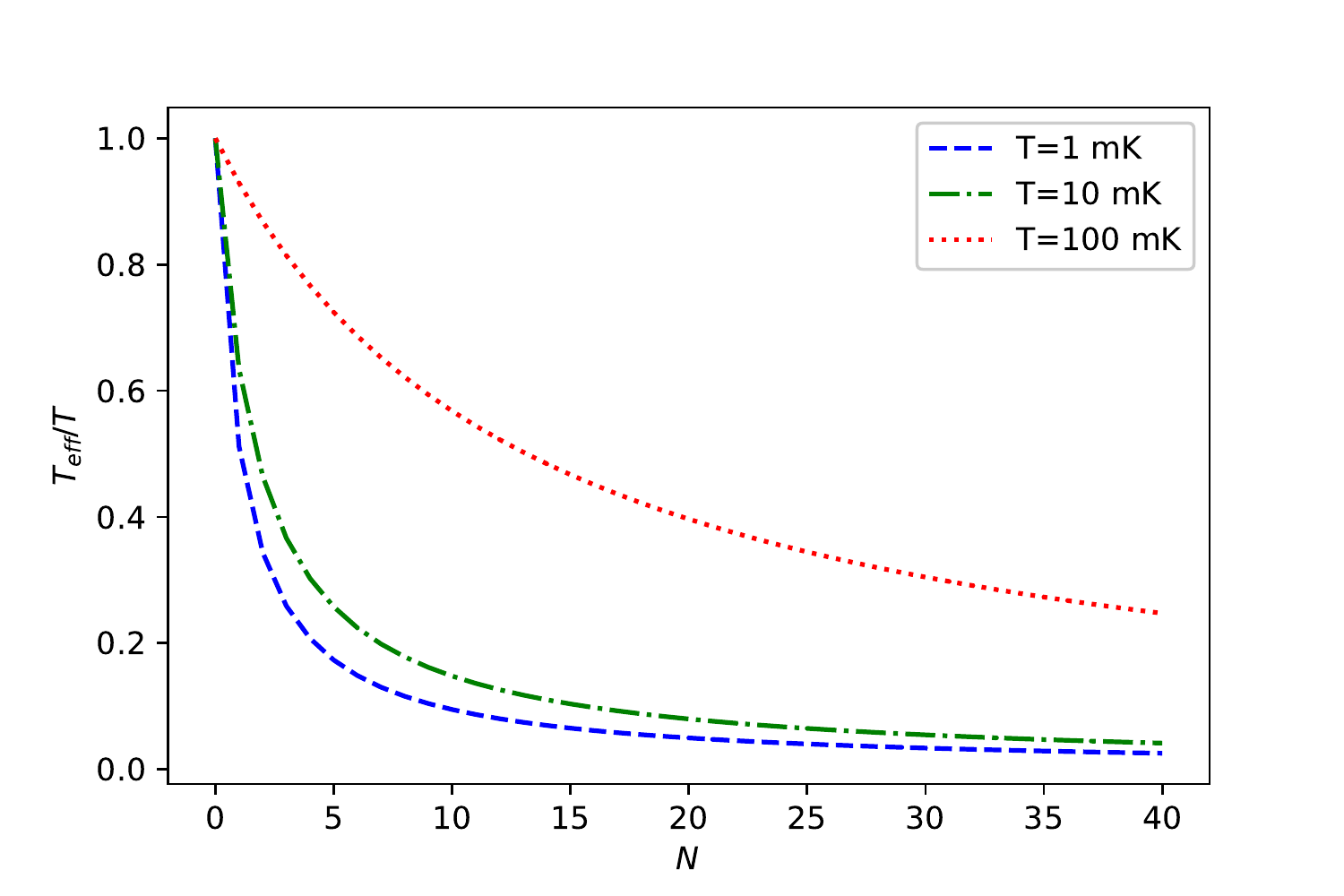}}
	\caption{\label{fig:ising-teff}Ratio of the effective $T_\text{eff}$ temperature of the central qubit to the environment temperature $T$ in a longitudinal ferromagnetic Ising spin-star model with $h=1$ GHz for (a) $N=6$ ancilla qubits at different interaction strengths $g$, (b) $g=-h$ at different number of ancilla qubits $N$.}
\end{figure} 

Taking the derivative of $\beta_{\text{eff}}$ with respect to the interaction strength $g$ here turns out to be insightful.
\begin{equation}
\frac{\partial \beta_{\text{eff}}}{\partial g} = \frac{-N\beta}{2h}(\tanh(\beta(h-g))+\tanh(\beta(h+g)))
\end{equation}
As $\tanh$ is a one-to-one odd function, setting the derivative to zero requires $h-g = -(h+g) = -h-g$, which is not satisfied for any value of $g$. Thus, $\beta_{\text{eff}}$ is a monotonic function of $g$ and evaluating the derivative for $g=0$ further shows that $\beta_{\text{eff}}$ is a monotonically decreasing function of $g$. For our purposes, this guarantees $\beta_{\text{eff}} > \beta$ when $g<0$, proving that our proposed setup manages to cool down the central qubit for ferromagnetic type interactions. Also, by monotonicity of $\beta_{\text{eff}}$ as a function of $g$, it keeps increasing while $g$ diverges towards $-\infty$, meaning that its limit at $-\infty$ is also its upper bound.
\begin{eqnarray}\label{eq:limitTeff}
\beta_{\text{max}} &=& \lim_{g \to -\infty} \beta_{\text{eff}} = \beta + \frac{N}{2h} \ln\left(\lim_{g \to -\infty} \frac{\cosh(\beta(h-g))}{\cosh(\beta(h+g))} \right) \nonumber \\
&=&  \beta + \frac{N}{2h} \ln\left(e^{2\beta h} \right) = (N+1)\beta \label{betamax}
\end{eqnarray}

Fig.~\ref{fig:ising-teff} shows the ratio of the effective temperature $T_\text{eff}$ to the environment temperature $T$ for different interaction strengths $g$ and number of ancilla qubits $N$. The asymptotic theoretical limit of the $T_\text{eff}$ in Eq.~(\ref{eq:limitTeff}) is approached faster with increasing $g$ in the low $T$ regime as shown in 
Fig.~\ref{fig:2a}. Fig.~\ref{fig:2b} suggest that, towards $g\sim -h$, reasonably large values of $N$ can achieve an order of magnitude cooling of the central qubit relative to typical environment temperatures in superconducting circuits ($<20$ mK).


\subsection{A simple refrigeration cycle to cool the central qubit and its efficiency\label{sec:refrigeratorCycle}}
To cool the central qubit,  we consider a cyclic transformation of the whole spin-star system in a single thermal environment.  
The cycle begins with uncoupled qubits ($g=0$) in thermal equilibrium at the environment temperature $T$. 

In the first step, the interaction is suddenly switched on so that there is no entropy change. At this stage, work is taken from the system, and there is no heat exchange with the environment. 

The interacting qubits (spin-star system) are left to thermalize to $T$ in the second step. While the spin-star system is in thermal equilibrium with the environment at $T$, the central spin is not. The effective temperature of the central qubit is given by Eq. (\ref{betaeff}). 

The third step consists of suddenly quenching the interaction ($g\rightarrow 0$) such that the state of the central qubit does not change. Under this assumption the transitions and the associated changes in $T_\text{eff}$ are negligible. In general, preservation of the initial state under a sudden perturbation requires that the switching on or off the interaction must be much faster than any characteristic time scale of the system, which is $1/2h$ for the central qubit. This condition is relaxed in our case, as the longitudinal Ising interactions (cf.~Eq.~(\ref{eq:model})) cannot cause any excitations in the initial thermal state before the quench. We can still introduce a bound on the perturbation time $\tau$. In practice, the qubits may not be uncoupled from the environment during the switching and hence we require $\tau<<1/\kappa$ where $\kappa$ is the relaxation (thermalization) time of the central qubit. Hence the central qubit remains cold at $T_\text{eff}$ for a duration of $\tau$. This
gives us a ``cooling window" in which the central qubit can be used
as a refrigerant to cool a many-qubit system, by the collisional route to thermalization, as described in Sec.~\ref{many} following Ref.~\cite{ourpaper}.
 
The fourth step is the thermalization of non-interacting central and ancilla qubits by the environment, bringing the whole system back to the beginning of the refrigeration cycle.
 
The cooling of the central qubit is performed with an efficiency given by 
\begin{equation}\label{effcy}
\varepsilon = \frac{E_s(\beta)-E_s(\beta_{\text{eff}})}
{W_{\text{cycle}}} =  \frac{h(\tanh(\beta_{\text{eff}} h)
-\tanh(\beta h))}{W_{\text{cycle}}}
\end{equation}
where $E_s(\beta)$ is the expectation of the bare system Hamiltonian at inverse temperature $\beta$ and $W_{\text{cycle}}$ is defined in Eq. (\ref{wcycle}) as the net work cost of turning on and off the Ising interactions. The interaction of the central qubit with the target many-body system at the end of the third step of the cycle does not affect the central qubit's cooling efficiency.

To calculate the efficiency defined in Eq.~(\ref{effcy}), we need the internal energy of the whole system at the end of each step of the cycle. The total energy is given by
\begin{equation}
E_{0}=-(N+1)h\tanh(\beta h)
\end{equation}
at the beginning of the cycle. After sudden quench by turning on the interaction, the state of the central qubit is preserved while 
the energy change is equal to the expectation of the interaction Hamiltonian at the initial state. As the state probability distribution of each qubit is independent, the total energy at the end of the first step $E_{1}$ can be calculated as 
\begin{equation}
E_{1}=-(N+1)h\tanh(\beta h) +\frac{gN(\cosh(2\beta h)-1)}{2\cosh^2(\beta h)}.
\end{equation}

We can calculate the energy of the interacting system in thermal equilibrium at the end of the second step by using the partition function in Eq.~(\ref{ztot}).
\begin{eqnarray}
&&E_{2} = -\frac{\partial \ln Z_{\text{tot}}}{\partial \beta} \nonumber \\
&&= \frac{-1}{e^{-\beta h}\cosh^N(\beta(g+h)) + e^{\beta h}\cosh^N(\beta(h-g))} \nonumber \\
&&\times(e^{-\beta h}\cosh^{N-1}(\beta(g+h))(N(g+h)\sinh(\beta(g+h))\nonumber\\
&&-h\cosh(\beta(g+h)))+e^{\beta h}\cosh^{N-1}(\beta(h-g))\nonumber\\
&&\times(\beta(h-g))(N(h-g)\sinh(\beta(h-g))+h\cosh(\beta(h-g))))\nonumber\\
&&\label{etot}
\end{eqnarray}

Finally, we can calculate the total energy of the system, $E_{3}$ after turning off the interaction at the end of the third step by calculating the expectation of the interaction Hamiltonian and subtracting it from $E_{2}$. 
\begin{eqnarray}
&&<\hat{H}_{\text{int}}> = \frac{-g}{\beta}\frac{\partial \ln Z_{\text{tot}}}{\partial g}\nonumber\\
&&=\frac{-gN}{e^{-\beta h}\cosh^N(\beta(g+h)) + e^{\beta h}\cosh^N(\beta(h-g))}\nonumber\\
&&\times(e^{-\beta h}\cosh^{N-1}(\beta(g+h))\sinh(\beta(g+h))\nonumber \\
&&-e^{\beta h}\cosh^{N-1}(\beta(h-g))\sinh(\beta(h-g)))\\
&&E_{3} = E_{2} - <\hat{H}_{\text{int}}> \label{eq2}
\end{eqnarray}

As Eqs.~(\ref{etot}) and (\ref{eq2}) are fairly long, we are not going to write down the explicit expression for the total work in a cycle and restrict ourselves to express it in terms of the energies at different stages of the cycle. 
\begin{eqnarray}
W_{\text{cycle}} &=& W_1 + W_2 = (E_{1}-E_{0})+(E_{3}-E_{2}) \nonumber\\
&=& \frac{gN(\cosh(2\beta h)-1)}{2\cosh^2(\beta h)} - <\hat{H}_{\text{int}}> \label{wcycle}
\end{eqnarray}


\begin{figure}[H]
	\centering
	\subfloat[\label{fig:3a}]{\includegraphics[width=\linewidth]
		{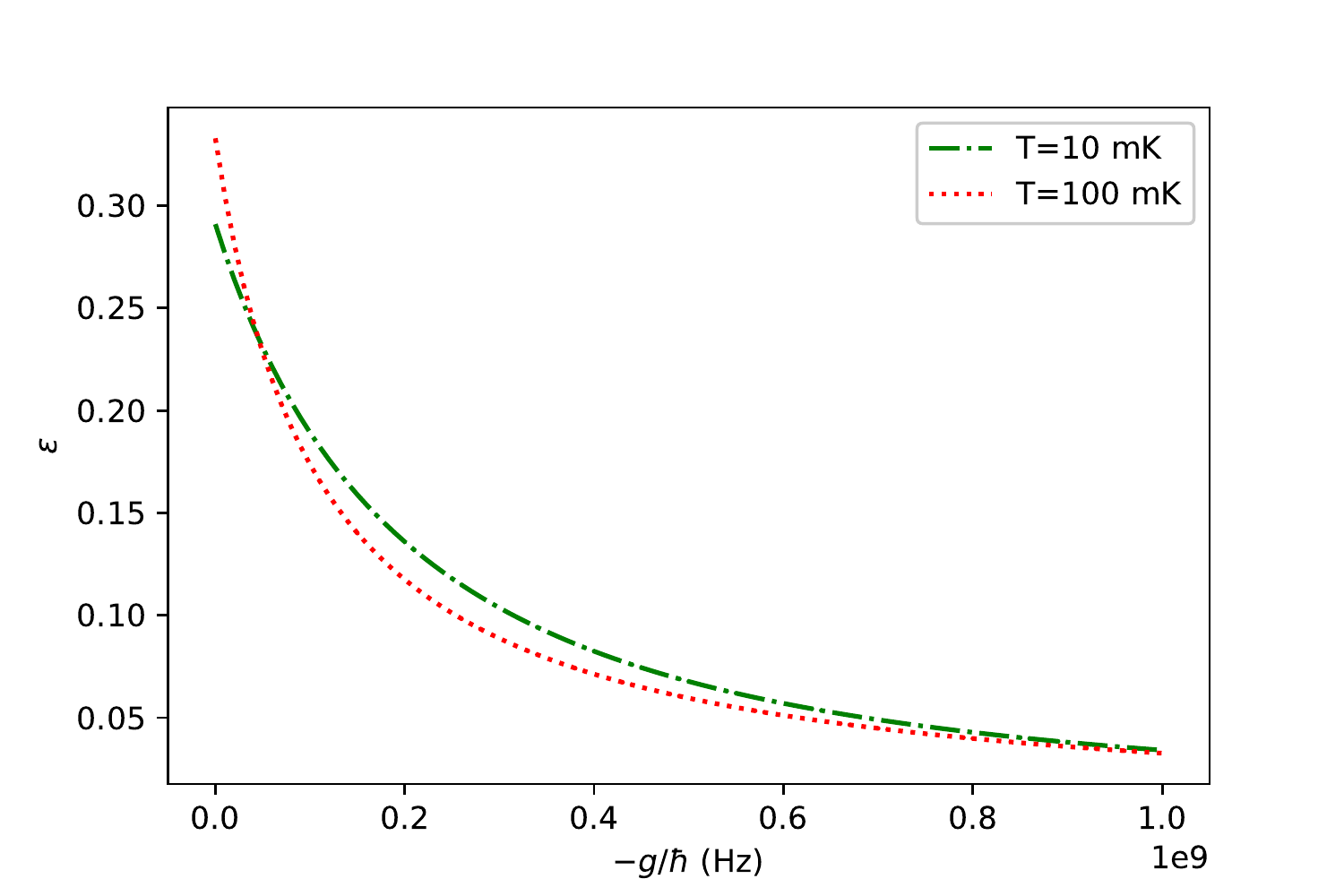}}
	\qquad
	\subfloat[\label{fig:3b}]{\includegraphics[width=\linewidth]
		{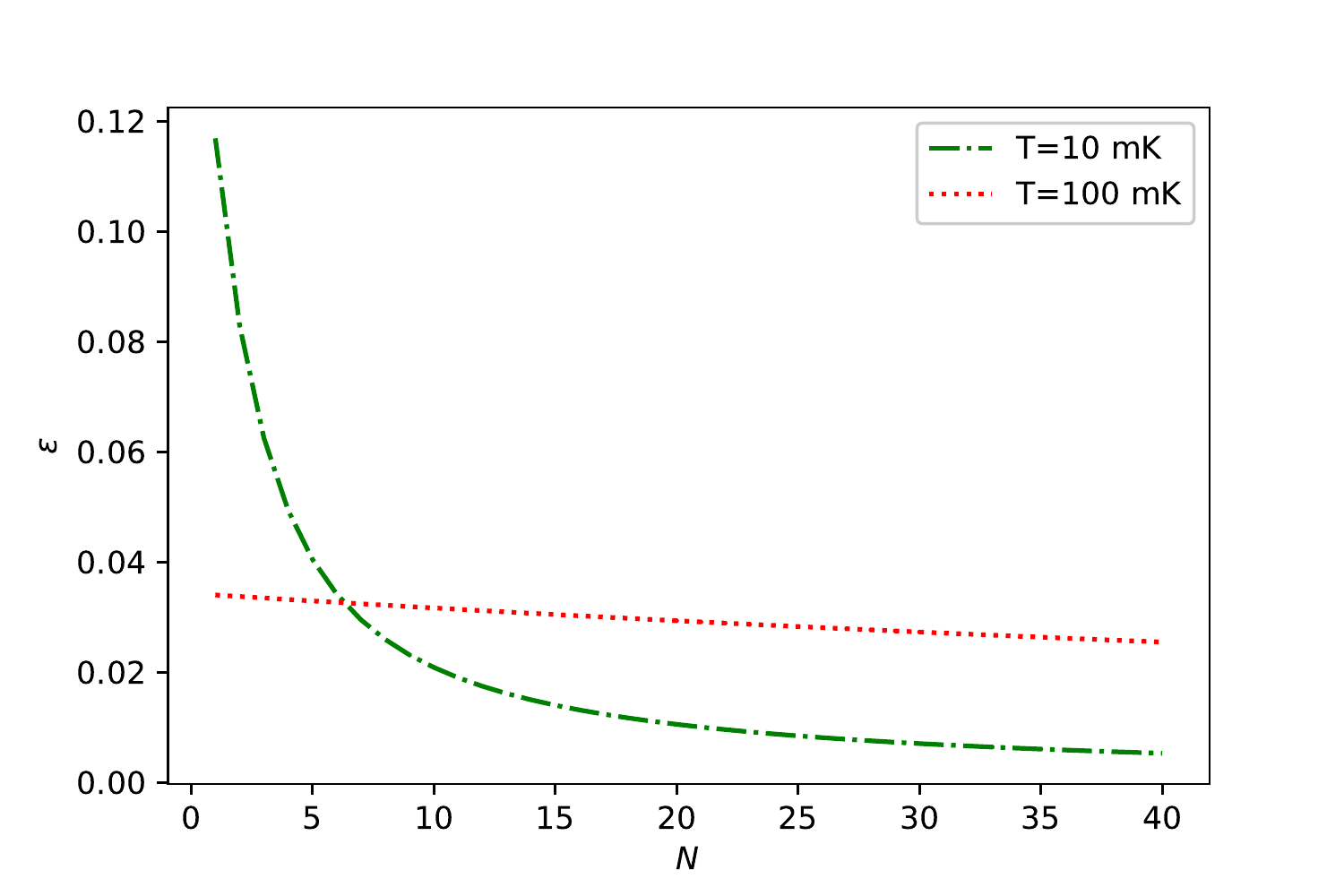}}
	\caption{\label{fig:ising-eff}Efficiency $\varepsilon$ of the 
		refrigeration cycle defined in Eq. (\ref{effcy}) as a function of (a) interaction strength $g$ with $N=6$ and (b) number of ancilla qubits $N$ with $g=-h$ at different environment temperatures $T$ and for $h=1$ GHz.}
\end{figure} 

The resulting efficiency with different number of ancilla qubits $N$ and different interaction strengths $g$ are plotted in Fig.~\ref{fig:ising-eff}. Fig.~\ref{fig:3a}
indicates that efficiency decreases with $g$. Comparing with Fig.~\ref{fig:2a},
we deduce that cooling to lower temperatures with increasing $g$ is not efficient.
Similar conclusion can be made for cooling by increasing $N$ after comparing Figs.~\ref{fig:2b} and~\ref{fig:3b}. An optimum strategy would be to use lower $g$ and $N$ values, relative to highest available ones, to cool to the target temperatures within acceptable efficiencies. For example, about an order of magnitude cooling can be achieved in typical superconducting qubit environment temperatures with $\sim 10\%$ efficiency for $g\sim -h/2$ and $N = 6$. In Sec.~\ref{ancilla-sec}, we will discuss exploiting the ancillae qubits to further increase the efficiency of the cooling cycle.

\subsection{Cooling of a many-body system with spin-star quantum refrigerators\label{many}}

\begin{figure}[t!]
	\includegraphics[width=\linewidth]{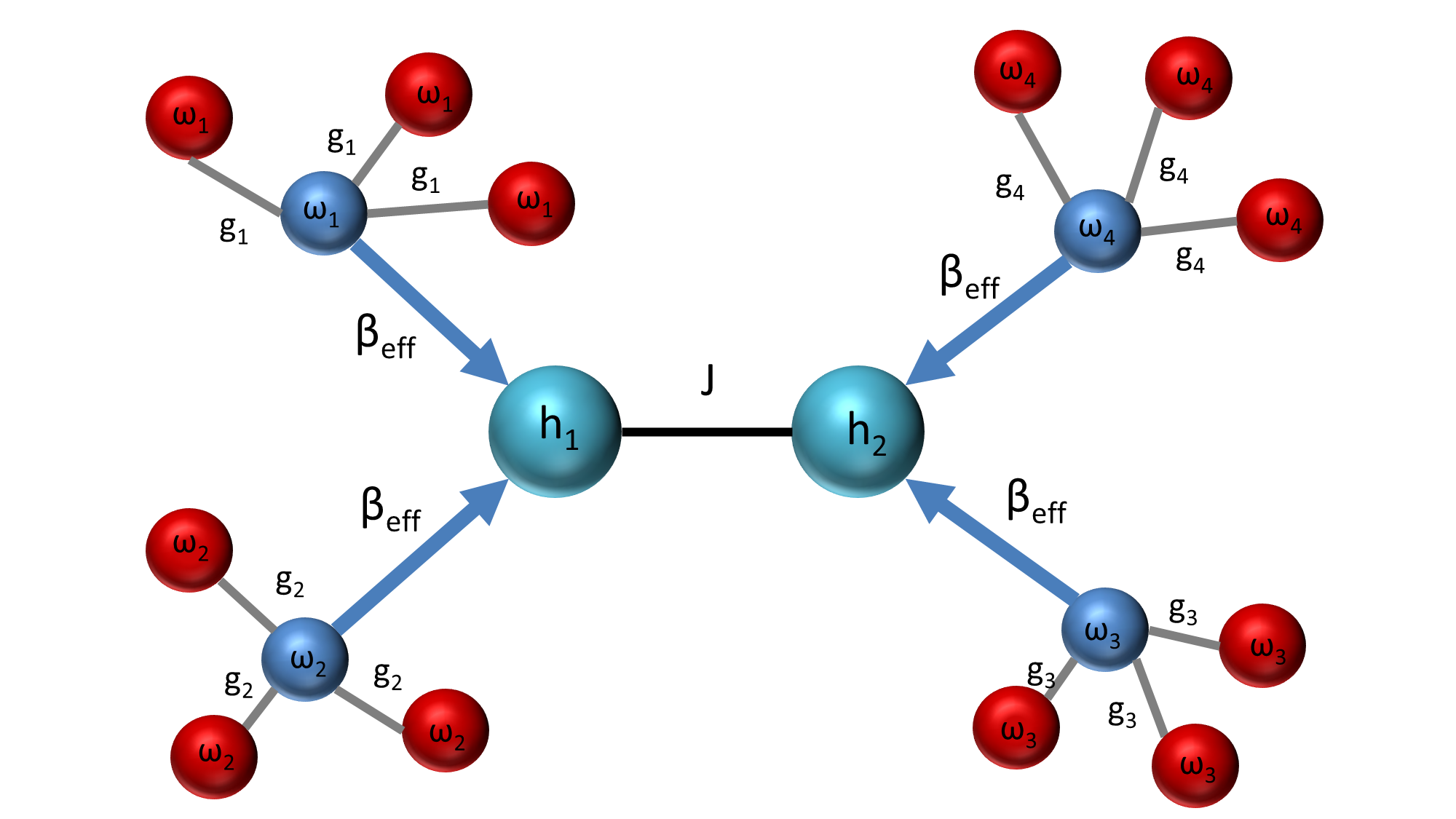}
	\caption{\label{fig:coll}Sketch of a Markovian collision model cooling a two-spin longitudinal Ising model described by the Hamilonian in Eq. (\ref{eq:targetModel}), with coupling strength $J$, using four spin-star quantum refrigerators labeled with $i=1..4$. Central qubits of the refrigerators are the refrigerants at effective inverse temperature $\beta_{\text{eff}}$. Central qubits are not resonant with the Ising model qubits, whose energy gaps are denoted by $h_1$ and $h_2$, instead, they are resonant with the transition frequencies $\omega_i$ ($i=1..4$) of the Ising model.
	Spin-star model has longitudinal and homogeneous couplings $g_i$.}
\end{figure}

We start the discussion of quantum many-body system cooling with a summary of the main results of Ref.~\cite{ourpaper}. 
The proposed scheme in Ref.~\cite{ourpaper} is for a general thermalization problem of a many-body system, consisting of interacting qubits. The system qubits make repeated collisions with a set of ``bath" qubits. The number of bath qubits depends on the number of transition frequencies of the many-body system. The scheme is suitable for cooling a small many-body system with a finite set of discrete eigenfrequencies in practice. Fig.~\ref{fig:coll} shows a case where a two-qubit system is thermalized with the collision model. The system Hamiltonian is taken to be a longitudinal Ising model 
\begin{eqnarray}\label{eq:targetModel}
H_\text{system}=\sum_{i=1}^2 h_i\sigma_{z,i}+J\sigma_{z,1}\sigma_{z,2},
\end{eqnarray}
which gives four transition frequencies $\omega_i$~\cite{ourpaper}. Here $h_i$ with $i=1,2$ are the resonant frequencies of the system qubits, and $J$ is the Ising coupling coefficients.
It is then sufficient to collide each system qubit with two-bath qubits at different $\omega_i$. In the present case, where our purpose is to cool down the system, the bath qubits are the central qubits coming out of the spin-star refrigerators at the third stage of the refrigeration cycle described in Sec.~\ref{sec:refrigeratorCycle}. Different spin-star refrigerators at different $h_i\equiv \omega_i /2$ should be adjusted to cool down their
central qubits to the same $T_\text{eff}$ by using different $g_i$ (cf.~Eq.~(\ref{betaeff})).

The derivation of the Markovian master equation in Lindblad form for a many-body collision model in Ref.~\cite{ourpaper} is based upon the set of standard assumptions of open quantum systems weakly coupled to large
reservoirs~\cite{breuer}. Starting with the Liouville-von Neumann equation for the system and environment coupling Hamiltonian $\hat{H}_I(t)$ in the interaction picture
\begin{eqnarray}
i\hbar\frac{\partial \rho}{\partial t} = [\hat{H}_I (t),\rho],
\end{eqnarray}
we integrate it over time. The zeroth and first-order solutions for the system-bath density matrix $\rho$ are plugged in the expression, and a second-order time-dependent perturbative equation is obtained. Assuming negligible change in the bath and system states, neglecting the system-bath entanglement, and applying the secular approximation to the resulting equation yield the well-known Markovian master equation for a large bath weakly coupled to a system for a long time~\cite{breuer}.

The assumption of a large bath coupled to the system for a long time is in sharp contrast with short-time collisions with a single two-level system. Nevertheless, we showed that one could ignore the finite time effects in the master equation under certain conditions~\cite{ourpaper}. First, the two-level system must be in resonance with one of the transition frequencies of the system. Second, the collisions must take longer time than the inverse of the transition frequency in question~\cite{ourpaper}. The resulting equation is given for a single two-level target system. It can still be generalized to systems with arbitrarily many energy levels by interpreting the master equation in the subspace spanned by states separated by the resonant frequency, as all the off-resonance terms are neglected by secular approximation. This leads to Lindblad dissipators in the following form for each collision
\begin{eqnarray}\label{eq:masterEqnManyBody}
D(\hat{\sigma}_{-},\hat{\sigma}_{+},\rho_s ) ~~&\propto&~~  (\rho_{gg}^{\text{bath}} (\hat{\sigma}_{-}\rho_{s}(t)\hat{\sigma}_{+} - \frac{1}{2}\{\hat{\sigma}_{+}\hat{\sigma}_{-},\rho_{s}(t)\}) \nonumber\\
&+& \rho_{ee}^{\text{bath}}(\hat{\sigma}_{+}\rho_{s}(t)\hat{\sigma}_{-} - \frac{1}{2}\{\hat{\sigma}_{-}\hat{\sigma}_{+},\rho_{s}(t)\})),\nonumber \\
&&
\label{me-tls}
\end{eqnarray}
where $\rho_{gg}^{\text{bath}}$ and $\rho_{ee}^{\text{bath}}$ are the ground and excited state populations of the colliding ``bath qubit" (central, refrigerant, qubit of the spin-star system) whose resonance frequency $\omega_i$ coincides with one of the transition frequencies of the system. The jump operators $\sigma_{\pm}$ are for a  system qubit. The density matrix of the many-qubit system is denoted by $\rho_s$.

Once the elimination of off-resonance terms is justified, the generalization to multiple transition frequencies is straightforward as the dissipators of collisions with different bath qubits are additive~\cite{ourpaper}. Each collision generates a term similar to 
Eq.~(\ref{me-tls}), responsible for transitions between two states separated by the bath qubit's frequency. The collision model depicted in Fig.~\ref{fig:coll} for the target many-body system with Hamiltonian given by Eq.~(\ref{eq:targetModel}) gives rise to the master equation
\begin{eqnarray}\label{me-mbs}
\frac{d}{dt}\rho_s &\propto& \sum_{i=1}^{2}\sum_{\omega_i}\left(\rho_{g,\omega_i}D(\hat{\sigma}_{-i}^{\omega_i},\hat{\sigma}_{+i}^{\omega_i},\rho_s)\right.\nonumber\\
&+&\left.\rho_{e,\omega_i}D(\hat{\sigma}_{+i}^{\omega_i},\hat{\sigma}_{-i}^{\omega_i},\rho_s)\right),
\end{eqnarray}
where $\hat{\sigma}_{\pm i}^{\omega_i}$ are the single-qubit transition operators for the $i$-th bath qubit at resonance frequency $\omega_i$~\cite{ourpaper}. $\rho_{g/e,\omega_i}$ are the ground/excited state populations of the bath qubits with resonance frequencies $\omega_i$.

The thermal state of the target multi-qubit system is the unique equilibrium point of the collisional master equation, Eq.~(\ref{me-mbs}), when the generated transitions connect all of the states of the system~\cite{ggl}. The Kubo-Martin-Schwinger (KMS) conditions for the resulting master equation show that the target system's equilibrium temperature is the same as that of the refrigerant qubits $T_\text{eff}$ ~\cite{breuer,ourpaper}. In summary, we conclude that a thermalizing master equation can describe the interaction of the central qubits with the many-qubit system for the system to evolve into a thermal equilibrium state with the refrigerant central qubits out of spin-star refrigerators.

Although not depicted in our sketch, the effect of the environment at a temperature $T > T_{\text{eff}}$ during the collisions also needs to be considered in a real application. Despite this setback, an appropriate choice of collision times and strengths can still bring the target system to equilibrium at a temperature $T_{\text{eff}}<T_{\text{eq}}<T$ as the dissipators due to the environment and the refrigerant qubits are additive.

\subsection{Final state of ancilla qubits and using them to enhance cooling efficiency\label{ancilla-sec}}
So far, we were only interested in the central qubit and traced out the ancilla qubits in all of our calculations. We also defined the efficiency in Eq.~(\ref{effcy}) by excluding the energy change in the ancilla qubits. This may be a drawback for our proposal for large numbers of ancilla qubits and cooling to very cold temperatures because the work cost of the cycle in Eq.~(\ref{wcycle}) is roughly proportional to the number of ancilla qubits while the energy extracted from the central qubit gets more or less saturated in very cold temperatures. As a workaround to this problem, we propose two possible uses of the ancilla qubits to increase the cooling efficiency. The first one is to use them in collisions with the many-qubit system for a cooperative effect and the second one is to use them in a heat engine cycle to help with the work required for running the spin-star refrigerators.

\subsubsection{Cooperative cooling with ancilla qubits}
\label{sec:coopCool}

\begin{figure}[t!]
\centering
	\subfloat[]{\includegraphics[width=\linewidth]{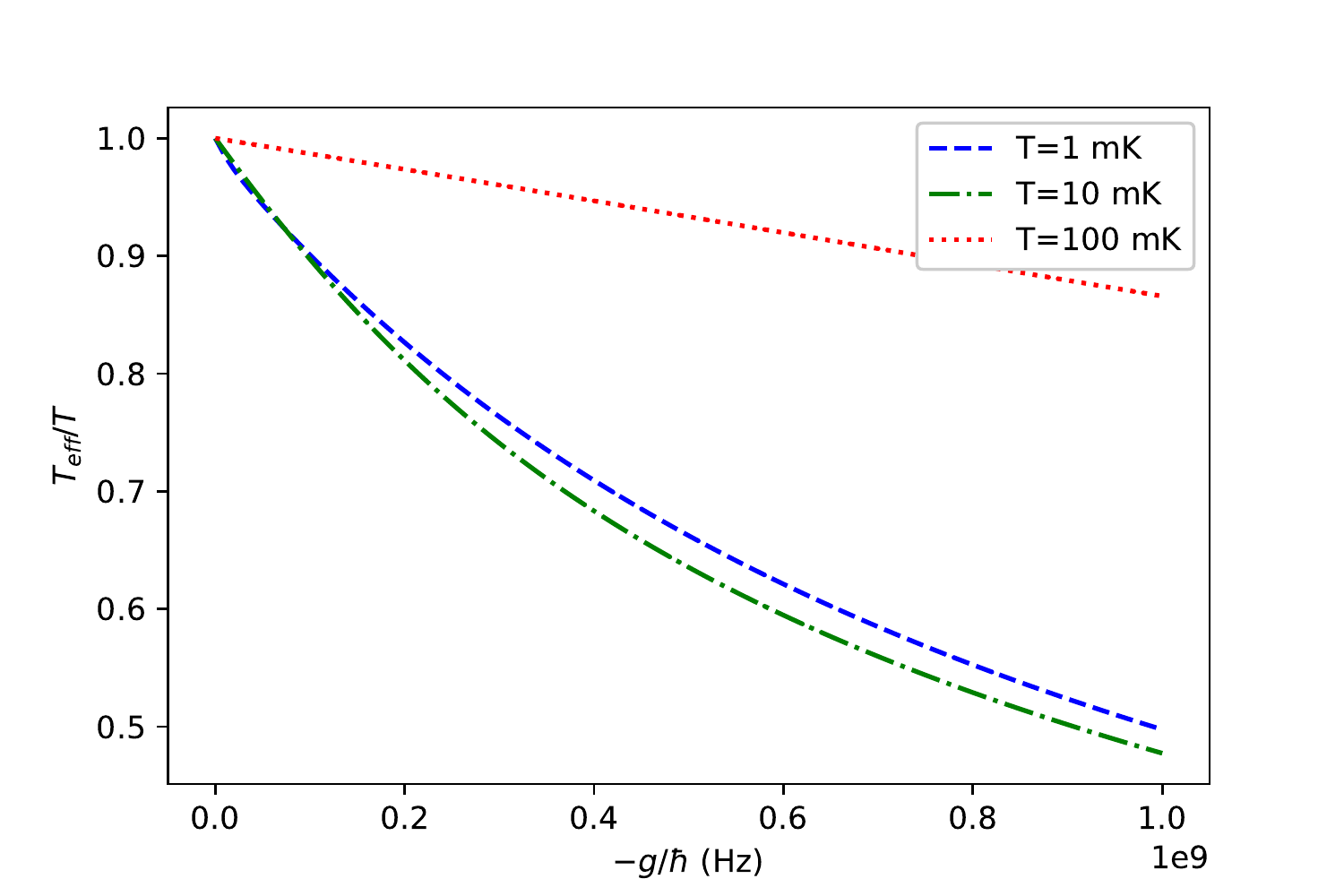}}
	\qquad
	\subfloat[]{\includegraphics[width=\linewidth]{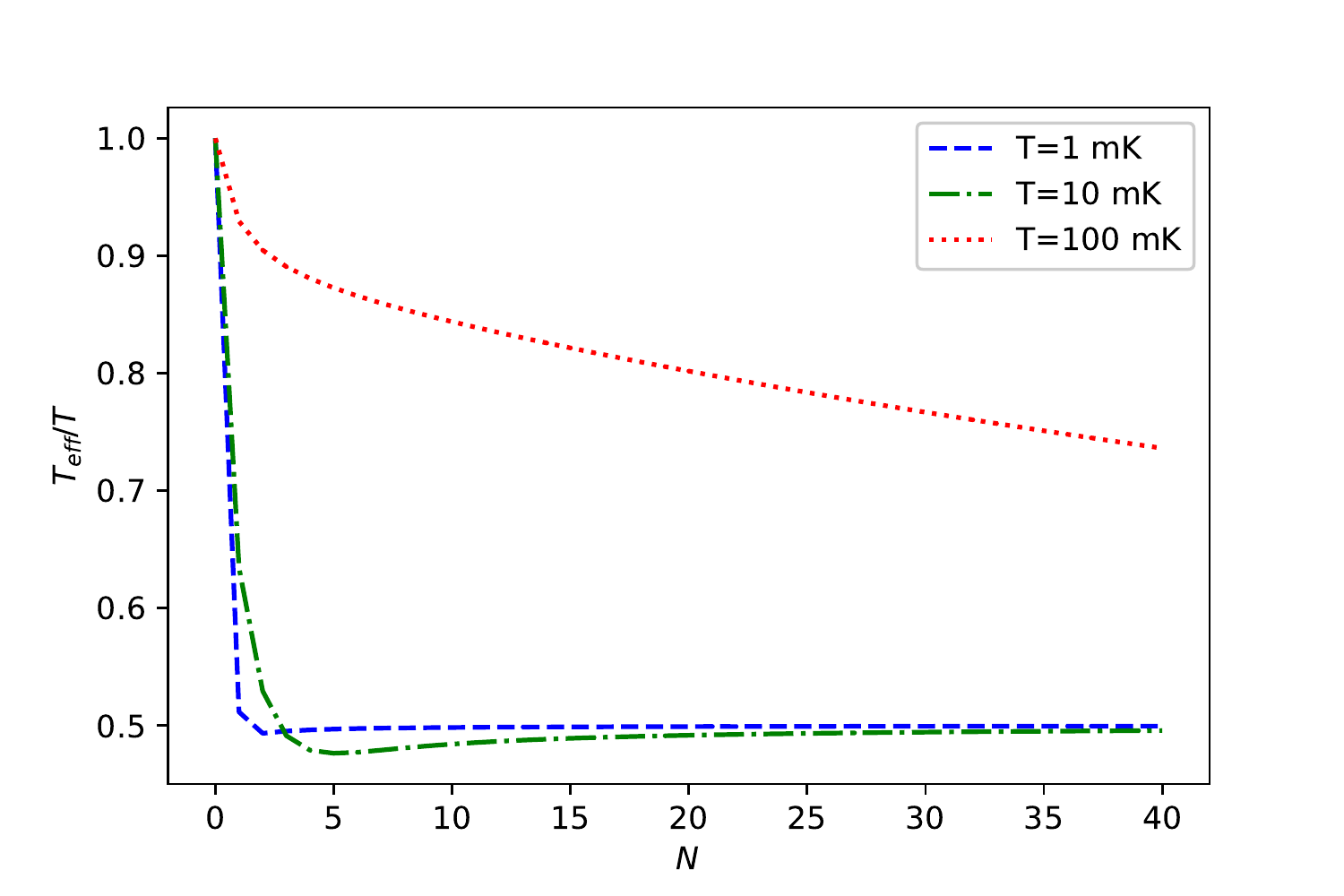}}
\caption{\label{fig:ising-total}Ratio of the effective temperature $T_\text{eff} = 1/k_B \beta_\text{eff,whole}$ of the whole spin-star system after turning off its Ising interactions defined in Eq. (\ref{beq}) to the environment temperature $T$ as a function of (a) interaction strength $g$ with $N=6$ and (b) number of ancilla $N$ qubits with $g=-h$. We take $h=1$ GHz.}
\end{figure}

Let's consider using the ancilla qubits together with the central qubit as the refrigerant of the spin-star refrigerator. The cooling dynamics of our scheme is described by a Markovian master equation in Eq.~(\ref{me-mbs}) with additive Lindblad dissipators for simultaneous collisions. When all the uncoupled qubits of the spin-star system in the third stage of the refrigerator cycle collide with a qubit of the target system simultaneously, the resulting
the master equation would be a straightforward generalization of Eq.~(\ref{me-mbs}).
The coefficients of two Lindblad dissipators in Eq.~(\ref{me-mbs})  responsible for heating and cooling become the sum of excited and ground state populations of the spin-star qubits, respectively. Accordingly,
the multi-qubit system relaxes to a thermal state at temperature $T_\text{eff,whole}$ which now depends on $N$. 

We can calculate $N_e$ and $N_g$ for a given set of spin-star qubits by using $N_e+N_g = N + 1$ and $N_e-N_g=\langle \hat{S}_z\rangle$ where 
$\hat{S}_z = \sum_{n=0}^{N} \hat{\sigma}_{z,n}$ and
\begin{eqnarray}
<\hat{S}_z > &=& \frac{-1}{\beta} \frac{\partial \ln Z_{\text{tot}}}{\partial h} = \frac{-2^N}{Z_{\text{tot}}}(e^{\beta h}\cosh^N (\beta(h-g))- \nonumber \\
&&e^{-\beta h}\cosh^N (\beta(h+g))+\nonumber \\
&&N(e^{-\beta h}\sinh(\beta(h+g))\cosh^{N-1} (\beta(h+g))+\nonumber \\
&&e^{\beta h}\sinh(\beta(h-g))\cosh^{N-1} (\beta(h-g)))).
\end{eqnarray}
$T_\text{eff,whole}$ is then given by 
\begin{equation}\label{beq}
\beta_{\text{eff,whole}} = \frac{1}{k_B T_{\text{eff,whole}}} = \frac{1}{2h} \ln \left( \frac{N+1-<\hat{S}_z>}{N+1+<\hat{S}_z>}\right).
\end{equation}

Cooling of the many-qubit system with transition
frequencies $\omega_i$ requires collisions
with sets of spin-star refrigerant qubits with $2h_i=\omega_i$.
Each spin-star cluster, associated with a different $\omega_i$, must be at the
same $T_\text{eff,whole} = 1/k_B \beta_\text{eff,whole}$, which can be
satisfied by using $g_i$. Under this condition, $T_\text{eff,whole}$ will
be the temperature of the multi-qubit system in a steady state due to the
repeated simultaneous collisions with the sets of the spin-star qubits.
Fig.~\ref{fig:ising-total} shows $T_\text{eff}$ for an example, where $\omega_i = 2$ GHz
so that $h_i \equiv h = 1$ GHz for a particular set of spin-star qubits.
For a target $T_\text{eff}$ one can determine the required $g_i \equiv g$
from Fig.~\ref{fig:ising-total}. Comparison of Fig.~\ref{fig:ising-teff} with Fig.~\ref{fig:ising-total} indicates that using only central qubits
as the refrigerants of the spin-star quantum refrigerators yields colder $T_\text{eff}$ for the many-body system. 

\begin{figure}[t!]
\centering
	\subfloat[\label{fig:6a}]{\includegraphics[width=\linewidth]{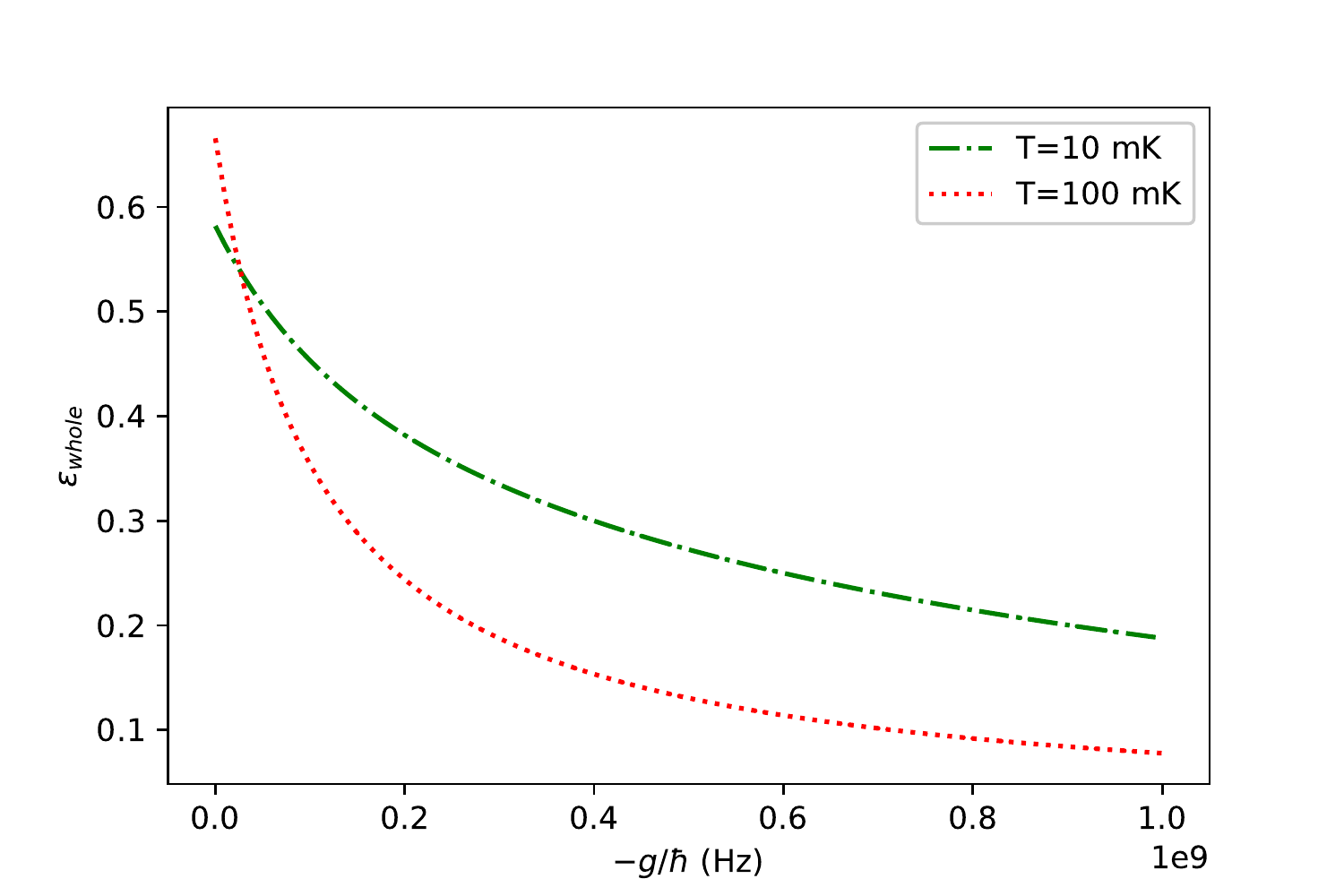}}
	\qquad
	\subfloat[\label{fig:6b}]{\includegraphics[width=\linewidth]{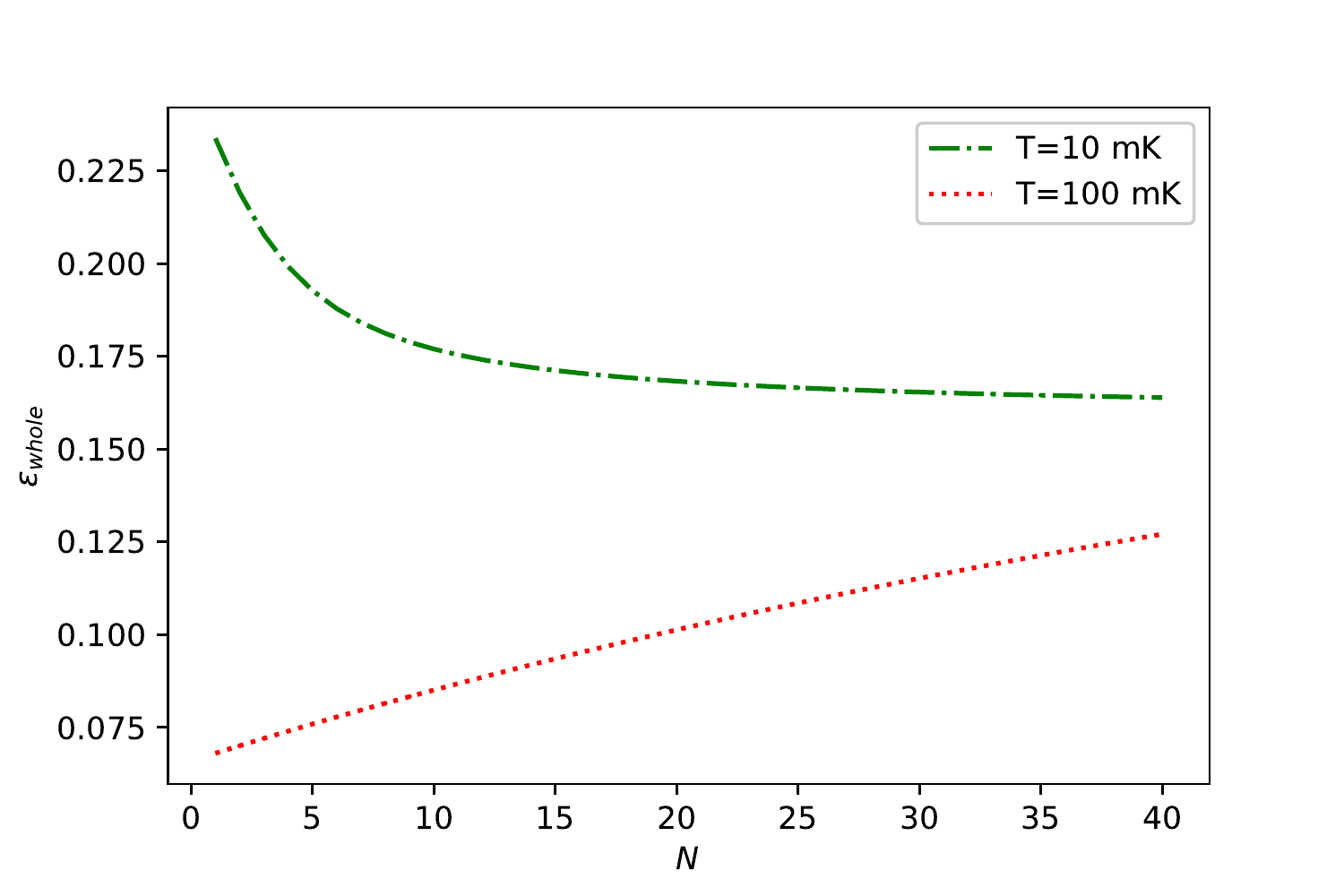}}
\caption{\label{fig:effcy-total}Efficiency $\varepsilon_{\text{whole}}$ defined in Eq. (\ref{ewhole}) as a function of (a) interaction strength $g$ with $N=6$ and (b) number of ancilla qubits $N$ with $g=-h$ at different environment temperatures $T$. We take
$h = 1$ GHz.}
\end{figure} 

As a concrete example of how limited this proposal is in terms of cooling the target many-body system, we observe from Fig.~\ref{fig:ising-total} that the ratio does not get significantly lower than $0.5$ for reasonable coupling strengths and unrealistically large numbers of ancilla qubits. We expect the relative advantage of using the whole spin star qubits should lie in the cooling efficiency. We define the efficiency of the cycle for cooperative cooling as 
\begin{equation}\label{ewhole}
\varepsilon_{\text{whole}} = \frac{E_0-E_{3}}{W_{\text{cycle}}}
\end{equation}
where the numerator is the total energy loss of the spin-star system instead of the energy loss of the central qubit only as in Eq. (\ref{effcy}) and the quantities $E_0$ and $E_3$ take the values calculated in Sec. \ref{sec:refrigeratorCycle}. The resulting efficiency with all the spin-star qubits for different $N$ and $g$ is plotted in Fig.~\ref{fig:effcy-total} which shows an anticipated increase in efficiency for all $N$ and $g$ compared to Eq. (\ref{effcy}). By a comparison with Fig.~\ref{fig:ising-eff}, the efficiency $\varepsilon_{\text{whole}}$ is several times higher than its counterpart $\varepsilon$ without the contribution of the ancilla qubits for most of the parameter choices. The increase of efficiency with the use of ancilla qubits is found to be particularly high in Fig.~\ref{fig:6b}, up to an order of magnitude for $T=10~\text{mK}$ which corresponds to the regime $h~ \text{\textasciitilde{}}~ k_B T/\hbar$ and high numbers of ancilla qubits. 

Based on our numerical results, we can conclude that the cooperative cooling with ancilla qubits always increases the efficiency but it significantly increases the minimum achievable effective temperature especially for high numbers of ancilla qubits compared to the case where only the central qubit is used for cooling of the target many-body system. However, this trade-off between achieving cooling to very cold temperatures and efficiency, which manifests itself as the dynamical third law of both classical~\cite{3rdlaw-2} and quantum~\cite{3rdlaw-1} thermodynamics, is the main challenge of all refrigeration schemes and it persists with our proposal. Also, cooperative cooling allows makes the thermalization of the target many-body system at the temperature $T_\text{eff,whole}$ faster and more robust against the inevitable effects of the environment on the many-body system.

To address the trade-off between reaching very low temperatures and refrigeration with high efficiency, we also consider discarding some of the ancilla qubits. For this purpose, we calculate the expectation of the operator defined as $\hat{S}'_z = \sum_{n=1}^{N} \hat{\sigma}_{z,n}$ by expressing the total spin-star Hamiltonian and its partition function as 
\begin{eqnarray}
\hat{H}_{\text{Ising}} &=& h_0~\hat{\sigma}_{z,0} + h_1\sum_{n=1}^{N}\hat{\sigma}_{z,n} + g~\hat{\sigma}_{z,0}\sum_{n=1}^{N}\hat{\sigma}_{z,n},\label{ising0} \\
Z_{\text{tot}} &=& 2^N (e^{-\beta h_0}\cosh^N(\beta(g+h_1))\nonumber \\
&&+e^{\beta h_0}\cosh^N(\beta(h_1-g))) .
\end{eqnarray}
We take $h_0=h_1=h$, which gives
\begin{eqnarray}
<\hat{S}'_z> &=& \frac{-1}{\beta} \frac{\partial \ln Z_{\text{tot}}}{\partial h_1} = \frac{-2^N N}{Z_{\text{tot}}}(e^{-\beta h}\sinh(\beta(h+g))\nonumber \\
&&\cosh^{N-1}(\beta(h+g))+e^{\beta h}\sinh(\beta(h-g)) \nonumber \\
&&\cosh^{N-1} (\beta(h-g))).
\end{eqnarray}

As the spin-star Hamiltonian is symmetric with respect to permutations of ancilla qubits, all of the ancilla qubits have the same ground and excited populations, so that we can calculate the effective temperature of ancilla spins similarly to Eq.~(\ref{beq}) as
\begin{equation}
\beta_{\text{eff,ancilla}} = \frac{1}{k_B T_{\text{eff,ancilla}}} = \frac{1}{2h} \ln \left( \frac{1-\frac{<\hat{S}'_z>}{N}}{1+\frac{<\hat{S}'_z>}{N}}\right). \label{tenv}
\end{equation}

\begin{figure}[t!]
\centering
	\subfloat[]{\includegraphics[width=\linewidth]{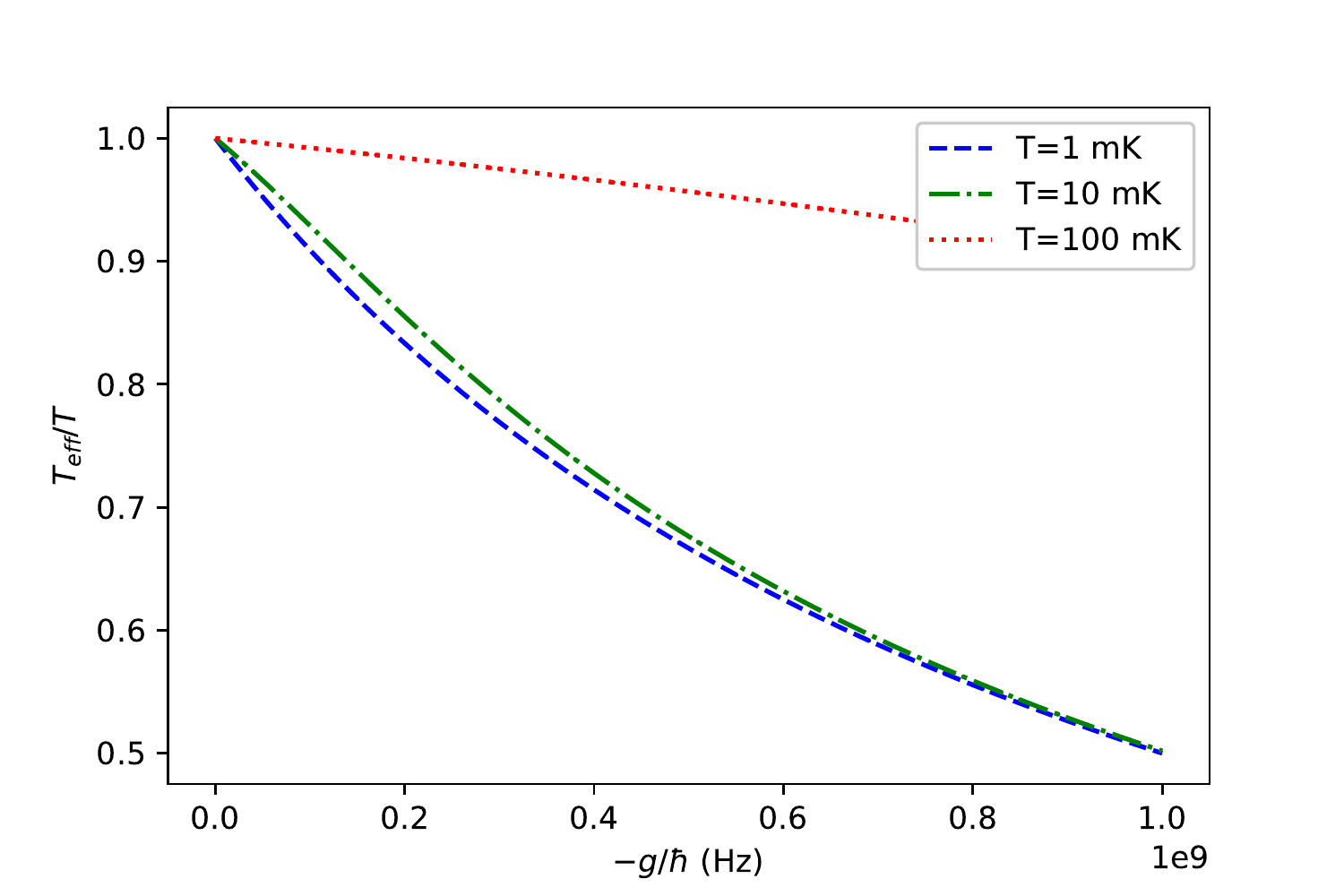}}
	\qquad
	\subfloat[]{\includegraphics[width=\linewidth]{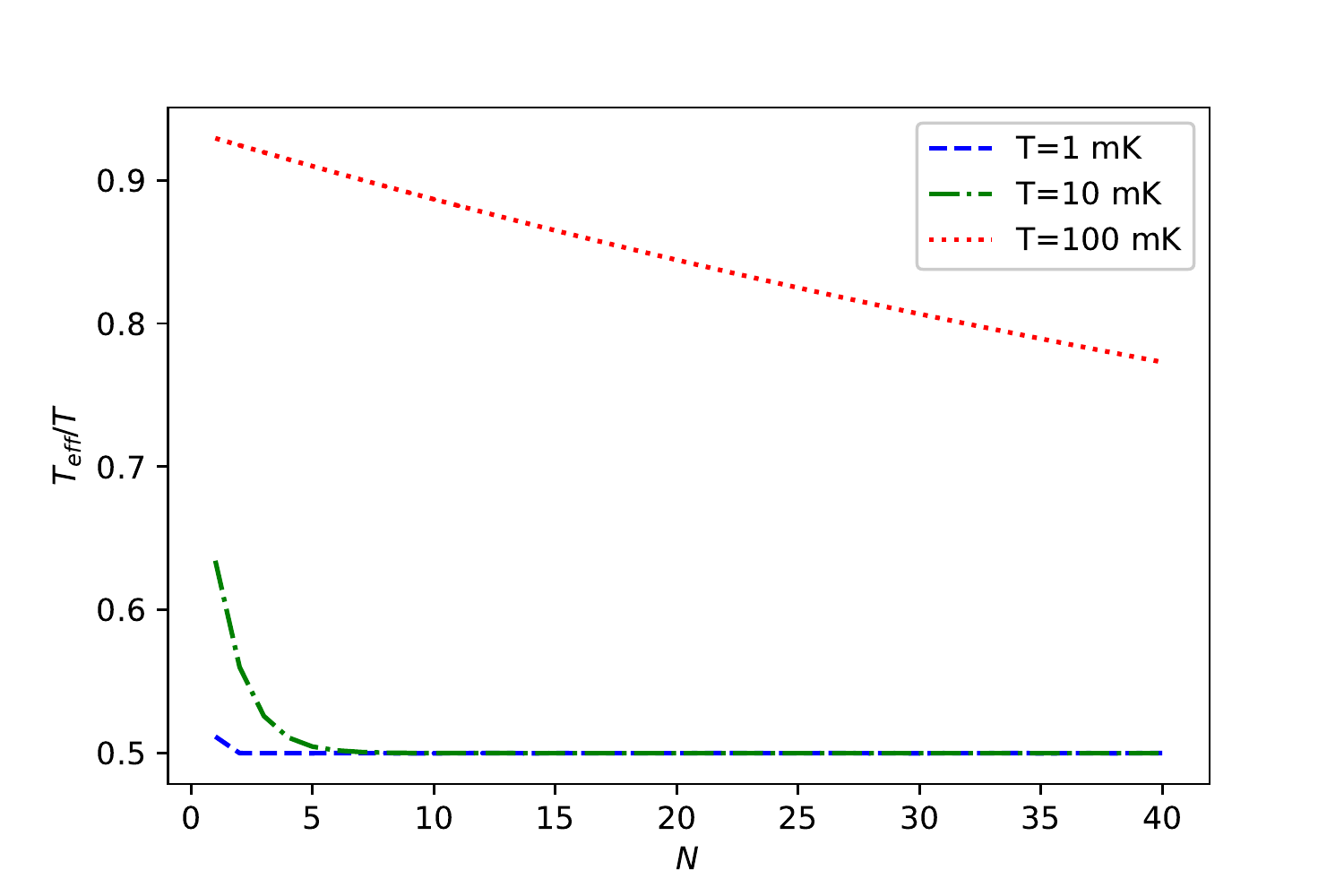}}
\caption{\label{fig:ising-env}Ratio of the effective temperature of the ancilla qubits $T_{\text{eff}} = 1/k_B \beta_{\text{eq,ancilla}}$ after turning off Ising interactions defined in Eq.~(\ref{tenv}) to the environment temperature $T$ as a function of (a) interaction strength $g$ with $N=6$ and (b) number of ancilla qubits $N$ with $g=-h$. We take
$h = 1$ GHz.}
\end{figure}

The resulting effective ancilla temperature is plotted in 
Fig.~\ref{fig:ising-env}. It is always higher than the center qubit effective temperature in Fig. \ref{fig:ising-teff} except for the trivial case of $N=1$ ancilla qubits. Therefore, the excited population of the ancilla qubits is always greater or equal to the excited population of the central qubit. 

Now, we can define an effective temperature of collective cooling when a number $n \leq N$ of the ancilla qubits are used as
\begin{equation}
\beta_{\text{eff,n}} = \frac{1}{k_B T_{\text{eff,n}}} = \frac{1}{2h} \ln \left( \frac{n+1-\frac{n<\hat{S}'_z>}{N}+\tanh(\beta_{\text{eff}}h)}{n+1+\frac{n<\hat{S}'_z>}{N}-\tanh(\beta_{\text{eff}}h)}\right).
\end{equation}
As we are able to see $|<\hat{S}'_z>/N|<\tanh(\beta_{\text{eff}}h)$ by comparing 
Figs.~\ref{fig:ising-teff} and~\ref{fig:ising-env}, we also have $\beta_{\text{eff}}>\beta_{\text{eff,n}}>\beta_{\text{eff,ancilla}}$. We can define the efficiency of the refrigeration cycle for the case of discarding some ancillae by ignoring the energy taken from these qubits, but the result is obvious: This efficiency would be between Eqs. (\ref{effcy}) and (\ref{ewhole}). 





\subsubsection{Ancilla qubits used as a cold bath for a quantum heat engine}
\label{sec:anc-engine}
Although using all qubits allows reasonable efficiency values for a specific temperature range, we propose another way of using the ancilla qubits to increase efficiency. As the center qubit's effective temperature gets lower with the increasing number of ancilla qubits while the effective temperature of ancilla qubits do not, we suggest that the center qubit can be used to cool down a many-body system to a very cold temperature. In contrast, the ancilla qubits can mimic a cold reservoir for an engine that would "recycle" some of the work spent in the refrigeration cycle after the Ising interaction of the spin-star system is turned off in a thermalized state, which corresponds to the interval between the third and fourth steps of the cycle described in Sec.~\ref{sec:refrigeratorCycle}. Similar to the many-body cooling discussed in the previous section, the interaction of the ancilla qubits with this engine must take place in a timescale much smaller than the relaxation time of the qubits to the environment temperature. For this proposal, the efficiency would depend on the type of engine in question, but we can give a reasonable definition of efficiency
\begin{equation}
\varepsilon_{\text{re}} = \frac{h(\tanh(\beta_{\text{eff,center}} h)-\tanh(\beta h))}{W_{\text{cycle}}-W_{\text{engine}}}
\end{equation}
based on the efficiency definition in Eq. (\ref{effcy}) without the contribution of the engine.

To gain insight into how large $W_{\text{engine}}$ can get, it is useful to calculate the effective temperature of the environment qubits after tracing out the center qubit by finding the ratio of the total ground and excited populations of the ancilla qubits. As all ancilla qubits are at the same effective temperature $\beta_{\text{eff,ancilla}}$, their collective effective temperature is also the same. This argument also applies to cases where some of the ancilla qubits are discarded. Fig.~\ref{fig:ising-env} shows the equilibrium temperature when all of the ancilla qubits in the spin-star system is used for collisions with the engine as its artificial cold reservoir.  The plot is somehow similar to Fig.~\ref{fig:ising-total} with center qubit included. 

Now that we have some qualitative results on the effective temperature of ancilla spins, we can make a more detailed comment on a possible engine working with the ancilla spins and its work production. As an analytically tractable \cite{ottorev} and experimentally realizable model \cite{iontrap}, we propose to use a quantum Otto engine using a harmonic oscillator as its working medium. For this engine, the environment would be the hot bath at the inverse temperature $\beta$ and the ancilla spins would be the cold bath at the inverse temperature $\beta_{\text{eq,ancilla}}$ using our previously proposed collision model~\cite{ourpaper}. 

As the thermalization of a system happens asymptotically with the number of collisions diverging to infinity, we assume that the number of ancilla spins $N$ is sufficiently large so that they are able to bring the harmonic oscillator to their effective temperature with negligible deviation. To summarize the quantum Otto cycle, the harmonic oscillator thermalized at the inverse temperature $\beta$ and the frequency $\omega_h$ is adiabatically driven to a lower frequency $\omega_c$, leading to a work extraction. Then, the harmonic oscillator is brought to the inverse temperature $\beta_{\text{eq,ancilla}}$ by collisions with ancilla spins, and it is driven back to the frequency $\omega_h$, taking some work from outside and completing the cycle. However, we cannot suppress the effects of the environment at the inverse temperature $\beta$ during the adiabatic strokes in an experimental realization of this engine, and we need to implement adiabatic strokes in short times so that the effect of the environment on these steps can be neglected, making these strokes strongly non-adiabatic and reducing the efficiency~\cite{nonad-otto1}. Another widely studied modification of this cycle is to introduce squeezing in the hot reservoir~\cite{ottosq1}, which is shown to exceed the Carnot efficiency~\cite{ottosq2} and even reach a unity efficiency for some choices of engine parameters~\cite{ottosq3}.

\section{Conclusion\label{sec:conclusion}}
In this work, we presented a way to cool down two-level systems using a finite number of ancilla spins and longitudinal ferromagnetic Ising interactions between center and ancilla spins. Our analytical calculations showed that the effective temperature of the center spin monotonically decreases with increasing magnitude of Ising interactions, and it asymptotically gets reduced by a factor of $N+1$, meaning that cooling the central spin by an order of magnitude with respect to its environment is not an unrealistic goal with currently available quantum technologies. We analyzed a simple refrigeration cycle in terms of its efficiency and proposed two different usages of ancilla spins after the refrigeration cycle to increase the efficiency. Based on our previous work \cite{ourpaper}, we also illustrated how our refrigerator for two-level systems could be a part of a many-body quantum system refrigerator, which is desirable in the context of quantum computation as pointed out in Refs.~\cite{ggl,metcalf2020} with different proposals of artificial environments for quantum many-body systems.

\appendix*

\section{Spin-star quantum refrigerator with Heisenberg model
interactions}

We argue here that the spin-star quantum heat engine model in Ref. \cite{epl-pce} can operate as a refrigerator as well, for an appropriate choice of parameters.  The model is different from our longitudinal Ising couplings due to the additional Heisenberg type transverse spin component interactions.  While the Ising interaction is suitable for an analytical study with classical statistical mechanics methods, we are not aware of any practical way of calculating the central qubit's density matrix with an arbitrary number of ancilla qubits. For this reason, we will restrict ourselves to show numerical results with a small number of ancilla qubits. 
The Hamiltonian of the whole system with Heisenberg interaction is given by
\begin{equation}
\hat{H}_{\text{Heisenberg}} = h\sum_{n=0}^{N}\hat{\sigma}_{z,n} + g\sum_{i=x,y,z}\sum_{n=1}^{N}\hat{\sigma}_{i,0}\hat{\sigma}_{i,n}. \label{hh}
\end{equation}
This choice corresponds to $\lambda=1$ in the notation of 
Ref.~\cite{epl-pce} for which the authors report a faster than linear increase of the ratio $T_{\text{eff}}/T_{\text{bath}}$ with respect to increasing number of environment qubits. Contrary to the assumptions of Ref.~\cite{epl-pce} and following our results for Ising interaction, we study the regime $g<0$ in Eq.~(\ref{hh}). 

\begin{figure}[H]
  \centering
  \includegraphics[width=\linewidth]{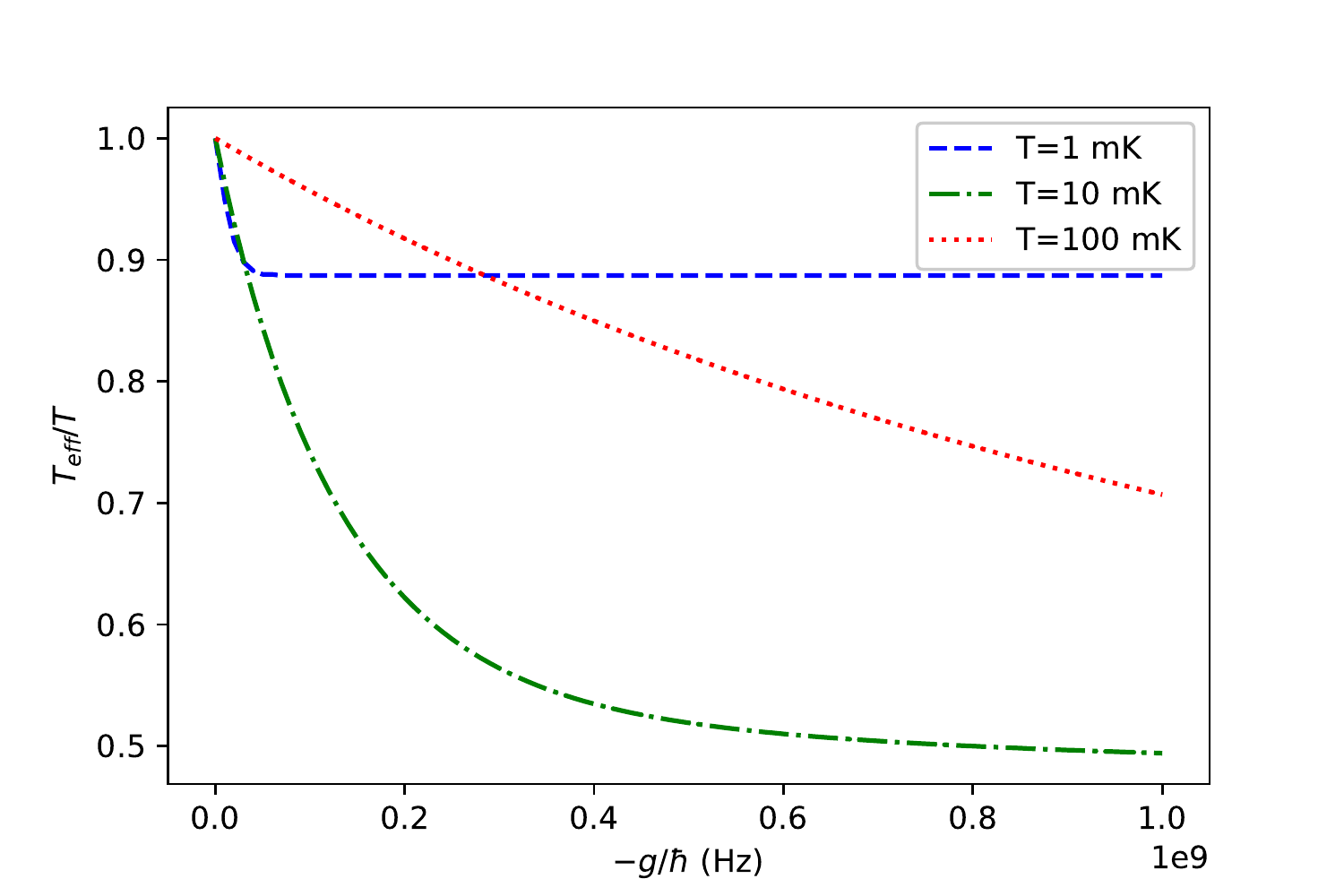}
\caption{\label{fig:hei-teff}Ratio of the effective $T_\text{eff}$ temperature of the central qubit to the environment temperature $T$ in a Heisenberg spin-star model for $N=6$ ancilla qubits at different interaction strengths $g$. We take $h=1$ GHz.}
\end{figure}

Fig.~\ref{fig:hei-teff} shows our numerical results for the ratio of effective temperature to the environment temperature for Heisenberg interaction. Our simulations for $2$ and $4$ ancilla qubits yielded very similar results, and we did not observe any significant change in the effective temperature of the central qubit. The Ising model is a better choice to significantly lower the effective temperature, especially with large numbers of ancilla qubits. The relatively poor cooling with Heisenberg interaction in our spin-star quantum refrigerator agrees with previous works reporting specific heat anomalies and non-zero excitation probability at zero temperature due to system-bath entanglement~\cite{entq,sha1,sha2}.

\begin{figure}[H]
  \includegraphics[width=\linewidth]{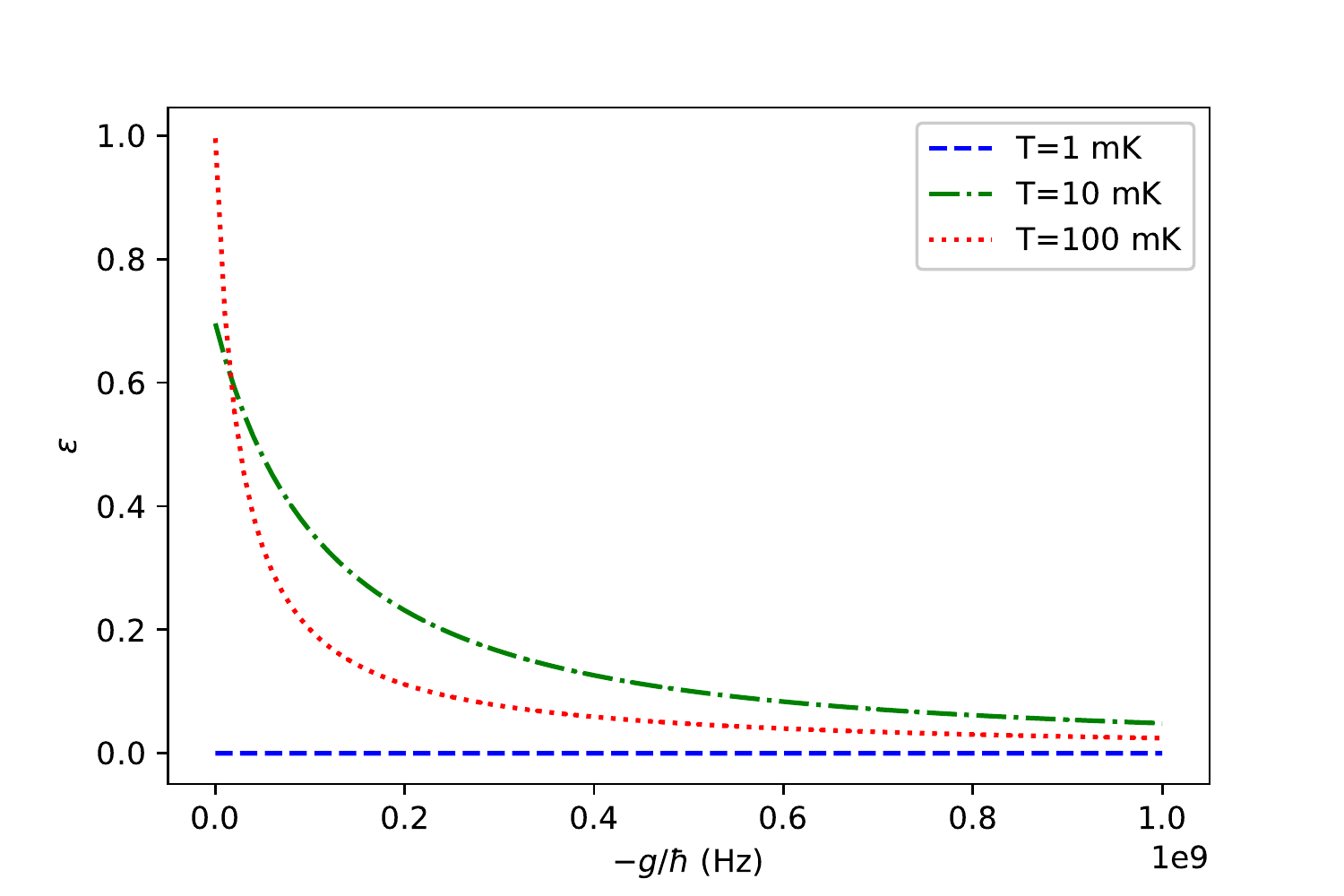}
\caption{\label{fig:hei-eff}Cooling efficiency $\varepsilon$ of a $(N+1)$-qubit Heisenberg spin-star model quantum refrigerator with $N=6$ ancilla qubits at different interaction strengths $g$.  We take $h=1$ GHz.}
\end{figure}

Fig.~\ref{fig:hei-eff} shows the efficiency of our refrigerator cycle described in Sec.~\ref{sec:refrigeratorCycle} with different temperatures and different interaction strengths for six ancilla qubits for Heisenberg interactions. Heisenberg interaction seems to be much more efficient at first sight; however, we need to emphasize that a poor refrigeration performance accompanies its high efficiency.


\bibliography{ising_refrigerator} 

\begin{thebibliography}{47}%
\makeatletter
\providecommand \@ifxundefined [1]{%
 \@ifx{#1\undefined}
}%
\providecommand \@ifnum [1]{%
 \ifnum #1\expandafter \@firstoftwo
 \else \expandafter \@secondoftwo
 \fi
}%
\providecommand \@ifx [1]{%
 \ifx #1\expandafter \@firstoftwo
 \else \expandafter \@secondoftwo
 \fi
}%
\providecommand \natexlab [1]{#1}%
\providecommand \enquote  [1]{``#1''}%
\providecommand \bibnamefont  [1]{#1}%
\providecommand \bibfnamefont [1]{#1}%
\providecommand \citenamefont [1]{#1}%
\providecommand \href@noop [0]{\@secondoftwo}%
\providecommand \href [0]{\begingroup \@sanitize@url \@href}%
\providecommand \@href[1]{\@@startlink{#1}\@@href}%
\providecommand \@@href[1]{\endgroup#1\@@endlink}%
\providecommand \@sanitize@url [0]{\catcode `\\12\catcode `\$12\catcode
  `\&12\catcode `\#12\catcode `\^12\catcode `\_12\catcode `\%12\relax}%
\providecommand \@@startlink[1]{}%
\providecommand \@@endlink[0]{}%
\providecommand \url  [0]{\begingroup\@sanitize@url \@url }%
\providecommand \@url [1]{\endgroup\@href {#1}{\urlprefix }}%
\providecommand \urlprefix  [0]{URL }%
\providecommand \Eprint [0]{\href }%
\providecommand \doibase [0]{https://doi.org/}%
\providecommand \selectlanguage [0]{\@gobble}%
\providecommand \bibinfo  [0]{\@secondoftwo}%
\providecommand \bibfield  [0]{\@secondoftwo}%
\providecommand \translation [1]{[#1]}%
\providecommand \BibitemOpen [0]{}%
\providecommand \bibitemStop [0]{}%
\providecommand \bibitemNoStop [0]{.\EOS\space}%
\providecommand \EOS [0]{\spacefactor3000\relax}%
\providecommand \BibitemShut  [1]{\csname bibitem#1\endcsname}%
\let\auto@bib@innerbib\@empty
\bibitem [{\citenamefont {Landi}\ and\ \citenamefont
  {Paternostro}(2020)}]{landi_irreversible_2020}%
  \BibitemOpen
  \bibfield  {author} {\bibinfo {author} {\bibfnamefont {G.~T.}\ \bibnamefont
  {Landi}}\ and\ \bibinfo {author} {\bibfnamefont {M.}~\bibnamefont
  {Paternostro}},\ }\bibfield  {title} {\bibinfo {title} {Irreversible entropy
  production, from quantum to classical},\ }\href
  {http://arxiv.org/abs/2009.07668} {\bibfield  {journal} {\bibinfo  {journal}
  {arXiv:2009.07668 [cond-mat, physics:quant-ph]}\ } (\bibinfo {year}
  {2020})},\ \bibinfo {note} {arXiv: 2009.07668}\BibitemShut {NoStop}%
\bibitem [{\citenamefont {Kosloff}(2013)}]{kosloff_quantum_2013}%
  \BibitemOpen
  \bibfield  {author} {\bibinfo {author} {\bibfnamefont {R.}~\bibnamefont
  {Kosloff}},\ }\bibfield  {title} {\bibinfo {title} {Quantum {Thermodynamics}:
  {A} {Dynamical} {Viewpoint}},\ }\href {https://doi.org/10.3390/e15062100}
  {\bibfield  {journal} {\bibinfo  {journal} {Entropy}\ }\textbf {\bibinfo
  {volume} {15}},\ \bibinfo {pages} {2100} (\bibinfo {year} {2013})},\ \bibinfo
  {note} {number: 6 Publisher: Multidisciplinary Digital Publishing
  Institute}\BibitemShut {NoStop}%
\bibitem [{\citenamefont {Partovi}(1989)}]{partovi_quantum_1989}%
  \BibitemOpen
  \bibfield  {author} {\bibinfo {author} {\bibfnamefont {M.~H.}\ \bibnamefont
  {Partovi}},\ }\bibfield  {title} {\bibinfo {title} {Quantum thermodynamics},\
  }\href {https://doi.org/10.1016/0375-9601(89)90221-1} {\bibfield  {journal}
  {\bibinfo  {journal} {Physics Letters A}\ }\textbf {\bibinfo {volume}
  {137}},\ \bibinfo {pages} {440} (\bibinfo {year} {1989})}\BibitemShut
  {NoStop}%
\bibitem [{\citenamefont {{\"O}zdemir}\ and\ \citenamefont
  {M{\"u}stecapl{\i}o{\u g}lu}(2020)}]{ozdemir_quantum_2020}%
  \BibitemOpen
  \bibfield  {author} {\bibinfo {author} {\bibfnamefont {A.~T.}\ \bibnamefont
  {{\"O}zdemir}}\ and\ \bibinfo {author} {\bibfnamefont {{\"O}.~E.}\
  \bibnamefont {M{\"u}stecapl{\i}o{\u g}lu}},\ }\bibfield  {title} {\bibinfo
  {title} {Quantum thermodynamics and quantum coherence engines},\ }\href
  {https://journals.tubitak.gov.tr/physics/abstract.htm?id=28103} {\bibfield
  {journal} {\bibinfo  {journal} {TURKISH JOURNAL OF PHYSICS}\ }\textbf
  {\bibinfo {volume} {44}},\ \bibinfo {pages} {404} (\bibinfo {year} {2020})},\
  \bibinfo {note} {publisher: The Scientific and Technological Research Council
  of Turkey}\BibitemShut {NoStop}%
\bibitem [{\citenamefont {Vinjanampathy}\ and\ \citenamefont
  {Anders}(2016)}]{vinjanampathy_quantum_2016}%
  \BibitemOpen
  \bibfield  {author} {\bibinfo {author} {\bibfnamefont {S.}~\bibnamefont
  {Vinjanampathy}}\ and\ \bibinfo {author} {\bibfnamefont {J.}~\bibnamefont
  {Anders}},\ }\bibfield  {title} {\bibinfo {title} {Quantum thermodynamics},\
  }\href {https://doi.org/10.1080/00107514.2016.1201896} {\bibfield  {journal}
  {\bibinfo  {journal} {Contemporary Physics}\ }\textbf {\bibinfo {volume}
  {57}},\ \bibinfo {pages} {545} (\bibinfo {year} {2016})},\ \bibinfo {note}
  {publisher: Taylor \& Francis \_eprint:
  https://doi.org/10.1080/00107514.2016.1201896}\BibitemShut {NoStop}%
\bibitem [{\citenamefont {Naseem}\ \emph {et~al.}(2020)\citenamefont {Naseem},
  \citenamefont {Misra},\ and\ \citenamefont {M{\"u}stecaplıo{\u
  g}lu}}]{naseem_two-body_2020}%
  \BibitemOpen
  \bibfield  {author} {\bibinfo {author} {\bibfnamefont {M.~T.}\ \bibnamefont
  {Naseem}}, \bibinfo {author} {\bibfnamefont {A.}~\bibnamefont {Misra}},\ and\
  \bibinfo {author} {\bibfnamefont {{\"O}.~E.}\ \bibnamefont
  {M{\"u}stecaplıo{\u g}lu}},\ }\bibfield  {title} {\bibinfo {title} {Two-body
  quantum absorption refrigerators with optomechanical-like interactions},\
  }\href {https://doi.org/10.1088/2058-9565/ab8d89} {\bibfield  {journal}
  {\bibinfo  {journal} {Quantum Science and Technology}\ }\textbf {\bibinfo
  {volume} {5}},\ \bibinfo {pages} {035006} (\bibinfo {year} {2020})},\
  \bibinfo {note} {publisher: IOP Publishing}\BibitemShut {NoStop}%
\bibitem [{\citenamefont {Abah}\ \emph {et~al.}(2020)\citenamefont {Abah},
  \citenamefont {Paternostro},\ and\ \citenamefont
  {Lutz}}]{abah_shortcut-adiabaticity_2020}%
  \BibitemOpen
  \bibfield  {author} {\bibinfo {author} {\bibfnamefont {O.}~\bibnamefont
  {Abah}}, \bibinfo {author} {\bibfnamefont {M.}~\bibnamefont {Paternostro}},\
  and\ \bibinfo {author} {\bibfnamefont {E.}~\bibnamefont {Lutz}},\ }\bibfield
  {title} {\bibinfo {title} {Shortcut-to-adiabaticity quantum {Otto}
  refrigerator},\ }\href {https://doi.org/10.1103/PhysRevResearch.2.023120}
  {\bibfield  {journal} {\bibinfo  {journal} {Physical Review Research}\
  }\textbf {\bibinfo {volume} {2}},\ \bibinfo {pages} {023120} (\bibinfo {year}
  {2020})},\ \bibinfo {note} {publisher: American Physical Society}\BibitemShut
  {NoStop}%
\bibitem [{\citenamefont {Allahverdyan}\ \emph {et~al.}(2004)\citenamefont
  {Allahverdyan}, \citenamefont {Graci\`a},\ and\ \citenamefont
  {Nieuwenhuizen}}]{pulse}%
  \BibitemOpen
  \bibfield  {author} {\bibinfo {author} {\bibfnamefont {A.~E.}\ \bibnamefont
  {Allahverdyan}}, \bibinfo {author} {\bibfnamefont {R.~S.}\ \bibnamefont
  {Graci\`a}},\ and\ \bibinfo {author} {\bibfnamefont {T.~M.}\ \bibnamefont
  {Nieuwenhuizen}},\ }\bibfield  {title} {\bibinfo {title} {Bath-assisted
  cooling of spins},\ }\href {https://doi.org/10.1103/PhysRevLett.93.260404}
  {\bibfield  {journal} {\bibinfo  {journal} {Phys. Rev. Lett.}\ }\textbf
  {\bibinfo {volume} {93}},\ \bibinfo {pages} {260404} (\bibinfo {year}
  {2004})}\BibitemShut {NoStop}%
\bibitem [{\citenamefont {Gelbwaser-Klimovsky}\ \emph
  {et~al.}(2013)\citenamefont {Gelbwaser-Klimovsky}, \citenamefont {Alicki},\
  and\ \citenamefont {Kurizki}}]{drivenref}%
  \BibitemOpen
  \bibfield  {author} {\bibinfo {author} {\bibfnamefont {D.}~\bibnamefont
  {Gelbwaser-Klimovsky}}, \bibinfo {author} {\bibfnamefont {R.}~\bibnamefont
  {Alicki}},\ and\ \bibinfo {author} {\bibfnamefont {G.}~\bibnamefont
  {Kurizki}},\ }\bibfield  {title} {\bibinfo {title} {Minimal universal quantum
  heat machine},\ }\href {https://doi.org/10.1103/PhysRevE.87.012140}
  {\bibfield  {journal} {\bibinfo  {journal} {Phys. Rev. E}\ }\textbf {\bibinfo
  {volume} {87}},\ \bibinfo {pages} {012140} (\bibinfo {year}
  {2013})}\BibitemShut {NoStop}%
\bibitem [{\citenamefont {Erez}\ \emph {et~al.}(2008)\citenamefont {Erez},
  \citenamefont {Gordon}, \citenamefont {Nest},\ and\ \citenamefont
  {Kurizki}}]{meas}%
  \BibitemOpen
  \bibfield  {author} {\bibinfo {author} {\bibfnamefont {N.}~\bibnamefont
  {Erez}}, \bibinfo {author} {\bibfnamefont {G.}~\bibnamefont {Gordon}},
  \bibinfo {author} {\bibfnamefont {M.}~\bibnamefont {Nest}},\ and\ \bibinfo
  {author} {\bibfnamefont {G.}~\bibnamefont {Kurizki}},\ }\bibfield  {title}
  {\bibinfo {title} {Thermodynamic control by frequent quantum measurements},\
  }\href {https://doi.org/10.1038/nature06873} {\bibfield  {journal} {\bibinfo
  {journal} {Nature}\ }\textbf {\bibinfo {volume} {452}},\ \bibinfo {pages}
  {724} (\bibinfo {year} {2008})}\BibitemShut {NoStop}%
\bibitem [{\citenamefont {Boykin}\ \emph {et~al.}(2002)\citenamefont {Boykin},
  \citenamefont {Mor}, \citenamefont {Roychowdhury}, \citenamefont {Vatan},\
  and\ \citenamefont {Vrijen}}]{alg1}%
  \BibitemOpen
  \bibfield  {author} {\bibinfo {author} {\bibfnamefont {P.~O.}\ \bibnamefont
  {Boykin}}, \bibinfo {author} {\bibfnamefont {T.}~\bibnamefont {Mor}},
  \bibinfo {author} {\bibfnamefont {V.}~\bibnamefont {Roychowdhury}}, \bibinfo
  {author} {\bibfnamefont {F.}~\bibnamefont {Vatan}},\ and\ \bibinfo {author}
  {\bibfnamefont {R.}~\bibnamefont {Vrijen}},\ }\bibfield  {title} {\bibinfo
  {title} {Algorithmic cooling and scalable nmr quantum computers},\ }\href
  {https://doi.org/10.1073/pnas.241641898} {\bibfield  {journal} {\bibinfo
  {journal} {Proc. Natl. Acad. Sci. U.S.A.}\ }\textbf {\bibinfo {volume}
  {99}},\ \bibinfo {pages} {3388} (\bibinfo {year} {2002})}\BibitemShut
  {NoStop}%
\bibitem [{\citenamefont {Fernandez}\ \emph {et~al.}(2004)\citenamefont
  {Fernandez}, \citenamefont {Lloyd}, \citenamefont {Mor},\ and\ \citenamefont
  {Roychowdhury}}]{alg2}%
  \BibitemOpen
  \bibfield  {author} {\bibinfo {author} {\bibfnamefont {J.~M.}\ \bibnamefont
  {Fernandez}}, \bibinfo {author} {\bibfnamefont {S.}~\bibnamefont {Lloyd}},
  \bibinfo {author} {\bibfnamefont {T.}~\bibnamefont {Mor}},\ and\ \bibinfo
  {author} {\bibfnamefont {V.}~\bibnamefont {Roychowdhury}},\ }\bibfield
  {title} {\bibinfo {title} {Algorithmic cooling of spins: A practicable method
  for increasing polarization},\ }\href
  {https://doi.org/10.1142/S0219749904000419} {\bibfield  {journal} {\bibinfo
  {journal} {Int. J. Quantum Inform.}\ }\textbf {\bibinfo {volume} {2}},\
  \bibinfo {pages} {461} (\bibinfo {year} {2004})}\BibitemShut {NoStop}%
\bibitem [{\citenamefont {Elias}\ \emph {et~al.}(2011)\citenamefont {Elias},
  \citenamefont {Mor},\ and\ \citenamefont {Weinstein}}]{alg3}%
  \BibitemOpen
  \bibfield  {author} {\bibinfo {author} {\bibfnamefont {Y.}~\bibnamefont
  {Elias}}, \bibinfo {author} {\bibfnamefont {T.}~\bibnamefont {Mor}},\ and\
  \bibinfo {author} {\bibfnamefont {Y.}~\bibnamefont {Weinstein}},\ }\bibfield
  {title} {\bibinfo {title} {Semioptimal practicable algorithmic cooling},\
  }\href {https://doi.org/10.1103/PhysRevA.83.042340} {\bibfield  {journal}
  {\bibinfo  {journal} {Phys. Rev. A}\ }\textbf {\bibinfo {volume} {83}},\
  \bibinfo {pages} {042340} (\bibinfo {year} {2011})}\BibitemShut {NoStop}%
\bibitem [{\citenamefont {K{\"o}se}\ \emph {et~al.}(2019)\citenamefont
  {K{\"o}se}, \citenamefont {{\c C}akmak}, \citenamefont {Gen{\c c}ten},
  \citenamefont {Kominis},\ and\ \citenamefont {M{\"u}stecaplıo{\u
  g}lu}}]{kose_algorithmic_2019}%
  \BibitemOpen
  \bibfield  {author} {\bibinfo {author} {\bibfnamefont {E.}~\bibnamefont
  {K{\"o}se}}, \bibinfo {author} {\bibfnamefont {S.}~\bibnamefont {{\c
  C}akmak}}, \bibinfo {author} {\bibfnamefont {A.}~\bibnamefont {Gen{\c
  c}ten}}, \bibinfo {author} {\bibfnamefont {I.~K.}\ \bibnamefont {Kominis}},\
  and\ \bibinfo {author} {\bibfnamefont {{\"O}.~E.}\ \bibnamefont
  {M{\"u}stecaplıo{\u g}lu}},\ }\bibfield  {title} {\bibinfo {title}
  {Algorithmic quantum heat engines},\ }\href
  {https://doi.org/10.1103/PhysRevE.100.012109} {\bibfield  {journal} {\bibinfo
   {journal} {Physical Review E}\ }\textbf {\bibinfo {volume} {100}},\ \bibinfo
  {pages} {012109} (\bibinfo {year} {2019})},\ \bibinfo {note} {publisher:
  American Physical Society}\BibitemShut {NoStop}%
\bibitem [{\citenamefont {Linden}\ \emph {et~al.}(2010)\citenamefont {Linden},
  \citenamefont {Popescu},\ and\ \citenamefont {Skrzypczyk}}]{smallref}%
  \BibitemOpen
  \bibfield  {author} {\bibinfo {author} {\bibfnamefont {N.}~\bibnamefont
  {Linden}}, \bibinfo {author} {\bibfnamefont {S.}~\bibnamefont {Popescu}},\
  and\ \bibinfo {author} {\bibfnamefont {P.}~\bibnamefont {Skrzypczyk}},\
  }\bibfield  {title} {\bibinfo {title} {How small can thermal machines be? the
  smallest possible refrigerator},\ }\href
  {https://doi.org/10.1103/PhysRevLett.105.130401} {\bibfield  {journal}
  {\bibinfo  {journal} {Phys. Rev. Lett.}\ }\textbf {\bibinfo {volume} {105}},\
  \bibinfo {pages} {130401} (\bibinfo {year} {2010})}\BibitemShut {NoStop}%
\bibitem [{\citenamefont {Dillenschneider}\ and\ \citenamefont
  {Lutz}(2009)}]{lutz2009}%
  \BibitemOpen
  \bibfield  {author} {\bibinfo {author} {\bibfnamefont {R.}~\bibnamefont
  {Dillenschneider}}\ and\ \bibinfo {author} {\bibfnamefont {E.}~\bibnamefont
  {Lutz}},\ }\bibfield  {title} {\bibinfo {title} {Energetics of quantum
  correlations},\ }\href {https://doi.org/10.1209/0295-5075/88/50003}
  {\bibfield  {journal} {\bibinfo  {journal} {Europhys. Lett.}\ }\textbf
  {\bibinfo {volume} {88}},\ \bibinfo {pages} {50003} (\bibinfo {year}
  {2009})}\BibitemShut {NoStop}%
\bibitem [{\citenamefont {Da\u{g}}\ \emph {et~al.}(2019)\citenamefont
  {Da\u{g}}, \citenamefont {Niedenzu}, \citenamefont {{\"{O}}zayd{\i}n},
  \citenamefont {M\"{u}stecapl{\i}o\u{g}lu},\ and\ \citenamefont
  {Kurizki}}]{coh3}%
  \BibitemOpen
  \bibfield  {author} {\bibinfo {author} {\bibfnamefont {C.~B.}\ \bibnamefont
  {Da\u{g}}}, \bibinfo {author} {\bibfnamefont {W.}~\bibnamefont {Niedenzu}},
  \bibinfo {author} {\bibfnamefont {F.}~\bibnamefont {{\"{O}}zayd{\i}n}},
  \bibinfo {author} {\bibfnamefont {{\"{O}}.~E.}\ \bibnamefont
  {M\"{u}stecapl{\i}o\u{g}lu}},\ and\ \bibinfo {author} {\bibfnamefont
  {G.}~\bibnamefont {Kurizki}},\ }\bibfield  {title} {\bibinfo {title}
  {Temperature control in dissipative cavities by entangled dimers},\ }\href
  {https://doi.org/10.1021/acs.jpcc.8b11445} {\bibfield  {journal} {\bibinfo
  {journal} {J. Phys. Chem. C}\ }\textbf {\bibinfo {volume} {123}},\ \bibinfo
  {pages} {4035−4043} (\bibinfo {year} {2019})}\BibitemShut {NoStop}%
\bibitem [{\citenamefont {Ar{\i}soy}\ \emph {et~al.}(2019)\citenamefont
  {Ar{\i}soy}, \citenamefont {Campbell},\ and\ \citenamefont
  {M\"{u}stecapl{\i}o\u{g}lu}}]{ourpaper}%
  \BibitemOpen
  \bibfield  {author} {\bibinfo {author} {\bibfnamefont {O.}~\bibnamefont
  {Ar{\i}soy}}, \bibinfo {author} {\bibfnamefont {S.}~\bibnamefont
  {Campbell}},\ and\ \bibinfo {author} {\bibfnamefont {{\"{O}}.~E.}\
  \bibnamefont {M\"{u}stecapl{\i}o\u{g}lu}},\ }\bibfield  {title} {\bibinfo
  {title} {Thermalization of finite many-body systems by a collision model},\
  }\href {https://doi.org/doi:10.3390/e21121182} {\bibfield  {journal}
  {\bibinfo  {journal} {Entropy}\ }\textbf {\bibinfo {volume} {21(12)}},\
  \bibinfo {pages} {1182} (\bibinfo {year} {2019})}\BibitemShut {NoStop}%
\bibitem [{\citenamefont {Strutt~(3rd Baron~Rayleigh)}(1891)}]{rayleigh1891}%
  \BibitemOpen
  \bibfield  {author} {\bibinfo {author} {\bibfnamefont {J.~W.}\ \bibnamefont
  {Strutt~(3rd Baron~Rayleigh)}},\ }\bibfield  {title} {\bibinfo {title} {Liii.
  dynamical problems in illustration of the theory of gases},\ }\href
  {https://doi.org/10.1080/14786449108620207} {\bibfield  {journal} {\bibinfo
  {journal} {The London, Edinburgh, and Dublin Philosophical Magazine and
  Journal of Science}\ }\textbf {\bibinfo {volume} {32}},\ \bibinfo {pages}
  {424} (\bibinfo {year} {1891})}\BibitemShut {NoStop}%
\bibitem [{\citenamefont {Scully}\ and\ \citenamefont
  {Lamb}(1967)}]{scully_quantum_1967}%
  \BibitemOpen
  \bibfield  {author} {\bibinfo {author} {\bibfnamefont {M.~O.}\ \bibnamefont
  {Scully}}\ and\ \bibinfo {author} {\bibfnamefont {W.~E.}\ \bibnamefont
  {Lamb}},\ }\bibfield  {title} {\bibinfo {title} {Quantum {Theory} of an
  {Optical} {Maser}. {I}. {General} {Theory}},\ }\href
  {https://doi.org/10.1103/PhysRev.159.208} {\bibfield  {journal} {\bibinfo
  {journal} {Physical Review}\ }\textbf {\bibinfo {volume} {159}},\ \bibinfo
  {pages} {208} (\bibinfo {year} {1967})},\ \bibinfo {note} {publisher:
  American Physical Society}\BibitemShut {NoStop}%
\bibitem [{\citenamefont {\c{C}akmak}\ \emph {et~al.}(2017)\citenamefont
  {\c{C}akmak}, \citenamefont {Manatuly},\ and\ \citenamefont
  {M\"{u}stecapl{\i}o\u{g}lu}}]{coh1}%
  \BibitemOpen
  \bibfield  {author} {\bibinfo {author} {\bibfnamefont {B.}~\bibnamefont
  {\c{C}akmak}}, \bibinfo {author} {\bibfnamefont {A.}~\bibnamefont
  {Manatuly}},\ and\ \bibinfo {author} {\bibfnamefont {{\"{O}}.~E.}\
  \bibnamefont {M\"{u}stecapl{\i}o\u{g}lu}},\ }\bibfield  {title} {\bibinfo
  {title} {Thermal production, protection, and heat exchange of quantum
  coherences},\ }\href {https://doi.org/10.1103/PhysRevA.96.032117} {\bibfield
  {journal} {\bibinfo  {journal} {Phys. Rev. A}\ }\textbf {\bibinfo {volume}
  {96}},\ \bibinfo {pages} {032117} (\bibinfo {year} {2017})}\BibitemShut
  {NoStop}%
\bibitem [{\citenamefont {Da\u{g}}\ \emph {et~al.}(2016)\citenamefont
  {Da\u{g}}, \citenamefont {Niedenzu}, \citenamefont
  {M\"{u}stecapl{\i}o\u{g}lu},\ and\ \citenamefont {Kurizki}}]{coh2}%
  \BibitemOpen
  \bibfield  {author} {\bibinfo {author} {\bibfnamefont {C.~B.}\ \bibnamefont
  {Da\u{g}}}, \bibinfo {author} {\bibfnamefont {W.}~\bibnamefont {Niedenzu}},
  \bibinfo {author} {\bibfnamefont {{\"{O}}.~E.}\ \bibnamefont
  {M\"{u}stecapl{\i}o\u{g}lu}},\ and\ \bibinfo {author} {\bibfnamefont
  {G.}~\bibnamefont {Kurizki}},\ }\bibfield  {title} {\bibinfo {title}
  {Multiatom quantum coherences in micromasers as fuel for thermal and
  nonthermal machines},\ }\href {https://doi.org/10.3390/e18070244} {\bibfield
  {journal} {\bibinfo  {journal} {Entropy}\ }\textbf {\bibinfo {volume}
  {18(7)}},\ \bibinfo {pages} {244} (\bibinfo {year} {2016})}\BibitemShut
  {NoStop}%
\bibitem [{\citenamefont {T\"{u}rkpen\c{c}e}\ \emph {et~al.}(2017)\citenamefont
  {T\"{u}rkpen\c{c}e}, \citenamefont {Alt{\i}nta\c{s}}, \citenamefont
  {Paternostro},\ and\ \citenamefont {M\"{u}stecapl{\i}o\u{g}lu}}]{epl-pce}%
  \BibitemOpen
  \bibfield  {author} {\bibinfo {author} {\bibfnamefont {D.}~\bibnamefont
  {T\"{u}rkpen\c{c}e}}, \bibinfo {author} {\bibfnamefont {F.}~\bibnamefont
  {Alt{\i}nta\c{s}}}, \bibinfo {author} {\bibfnamefont {M.}~\bibnamefont
  {Paternostro}},\ and\ \bibinfo {author} {\bibfnamefont {{\"{O}}.~E.}\
  \bibnamefont {M\"{u}stecapl{\i}o\u{g}lu}},\ }\bibfield  {title} {\bibinfo
  {title} {A photonic carnot engine powered by a spin-star network},\ }\href
  {https://doi.org/10.1209/0295-5075/117/50002} {\bibfield  {journal} {\bibinfo
   {journal} {Europhys. Lett.}\ }\textbf {\bibinfo {volume} {117}},\ \bibinfo
  {pages} {50002} (\bibinfo {year} {2017})}\BibitemShut {NoStop}%
\bibitem [{\citenamefont {Cattaneo}\ \emph {et~al.}(2020)\citenamefont
  {Cattaneo}, \citenamefont {de~Chiara}, \citenamefont {Maniscalco},
  \citenamefont {Zambrini},\ and\ \citenamefont {Giorgi}}]{mark-coll}%
  \BibitemOpen
  \bibfield  {author} {\bibinfo {author} {\bibfnamefont {M.}~\bibnamefont
  {Cattaneo}}, \bibinfo {author} {\bibfnamefont {G.}~\bibnamefont {de~Chiara}},
  \bibinfo {author} {\bibfnamefont {S.}~\bibnamefont {Maniscalco}}, \bibinfo
  {author} {\bibfnamefont {R.}~\bibnamefont {Zambrini}},\ and\ \bibinfo
  {author} {\bibfnamefont {G.~L.}\ \bibnamefont {Giorgi}},\ }\bibfield  {title}
  {\bibinfo {title} {Collision models can efficiently simulate any multipartite
  markovian quantum dynamics},\ }\href {https://arxiv.org/abs/2010.13910}
  {\bibfield  {journal} {\bibinfo  {journal} {arXiv:2010.13910}\ } (\bibinfo
  {year} {2020})}\BibitemShut {NoStop}%
\bibitem [{\citenamefont {Ising}(1925)}]{ising}%
  \BibitemOpen
  \bibfield  {author} {\bibinfo {author} {\bibfnamefont {E.}~\bibnamefont
  {Ising}},\ }\bibfield  {title} {\bibinfo {title} {Beitrag zur theorie des
  ferromagnetismus},\ }\href {https://doi.org/10.1007/BF02980577} {\bibfield
  {journal} {\bibinfo  {journal} {Z. Physik}\ }\textbf {\bibinfo {volume}
  {31}},\ \bibinfo {pages} {253} (\bibinfo {year} {1925})}\BibitemShut
  {NoStop}%
\bibitem [{\citenamefont {Schliemann}\ \emph {et~al.}(2003)\citenamefont
  {Schliemann}, \citenamefont {Khaetskii},\ and\ \citenamefont {Loss}}]{qdot1}%
  \BibitemOpen
  \bibfield  {author} {\bibinfo {author} {\bibfnamefont {J.}~\bibnamefont
  {Schliemann}}, \bibinfo {author} {\bibfnamefont {A.}~\bibnamefont
  {Khaetskii}},\ and\ \bibinfo {author} {\bibfnamefont {D.}~\bibnamefont
  {Loss}},\ }\bibfield  {title} {\bibinfo {title} {Electron spin dynamics in
  quantum dots and related nanostructures due to hyperfine interaction with
  nuclei},\ }\href {https://doi.org/10.1088/0953-8984/15/50/r01} {\bibfield
  {journal} {\bibinfo  {journal} {J. Phys. Condens. Matter}\ }\textbf {\bibinfo
  {volume} {15}},\ \bibinfo {pages} {R1809} (\bibinfo {year}
  {2003})}\BibitemShut {NoStop}%
\bibitem [{\citenamefont {Kane}(1998)}]{qdot2}%
  \BibitemOpen
  \bibfield  {author} {\bibinfo {author} {\bibfnamefont {B.~E.}\ \bibnamefont
  {Kane}},\ }\bibfield  {title} {\bibinfo {title} {A silicon-based nuclear spin
  quantum computer},\ }\href {https://doi.org/10.1038/30156} {\bibfield
  {journal} {\bibinfo  {journal} {Nature}\ }\textbf {\bibinfo {volume} {393}},\
  \bibinfo {pages} {133} (\bibinfo {year} {1998})}\BibitemShut {NoStop}%
\bibitem [{\citenamefont {Breuer}\ \emph {et~al.}(2004)\citenamefont {Breuer},
  \citenamefont {Burgarth},\ and\ \citenamefont {Petruccione}}]{nm-ss1}%
  \BibitemOpen
  \bibfield  {author} {\bibinfo {author} {\bibfnamefont {H.~P.}\ \bibnamefont
  {Breuer}}, \bibinfo {author} {\bibfnamefont {D.}~\bibnamefont {Burgarth}},\
  and\ \bibinfo {author} {\bibfnamefont {F.}~\bibnamefont {Petruccione}},\
  }\bibfield  {title} {\bibinfo {title} {Non-markovian dynamics in a spin star
  system: Exact solution and approximation techniques},\ }\href
  {https://doi.org/10.1103/PhysRevB.70.045323} {\bibfield  {journal} {\bibinfo
  {journal} {Phys. Rev. B}\ }\textbf {\bibinfo {volume} {70}},\ \bibinfo
  {pages} {045323} (\bibinfo {year} {2004})}\BibitemShut {NoStop}%
\bibitem [{\citenamefont {Krovi}\ \emph {et~al.}(2007)\citenamefont {Krovi},
  \citenamefont {Oreshkov}, \citenamefont {Ryazanov},\ and\ \citenamefont
  {Lidar}}]{nm-ss2}%
  \BibitemOpen
  \bibfield  {author} {\bibinfo {author} {\bibfnamefont {H.}~\bibnamefont
  {Krovi}}, \bibinfo {author} {\bibfnamefont {O.}~\bibnamefont {Oreshkov}},
  \bibinfo {author} {\bibfnamefont {M.}~\bibnamefont {Ryazanov}},\ and\
  \bibinfo {author} {\bibfnamefont {D.~A.}\ \bibnamefont {Lidar}},\ }\bibfield
  {title} {\bibinfo {title} {Non-markovian dynamics of a qubit coupled to an
  ising spin bath},\ }\href {https://doi.org/10.1103/PhysRevA.76.052117}
  {\bibfield  {journal} {\bibinfo  {journal} {Phys. Rev. A}\ }\textbf {\bibinfo
  {volume} {76}},\ \bibinfo {pages} {052117} (\bibinfo {year}
  {2007})}\BibitemShut {NoStop}%
\bibitem [{\citenamefont {Wang}\ \emph {et~al.}(2013)\citenamefont {Wang},
  \citenamefont {Guo},\ and\ \citenamefont {Zhou}}]{nm-ss3}%
  \BibitemOpen
  \bibfield  {author} {\bibinfo {author} {\bibfnamefont {Z.~H.}\ \bibnamefont
  {Wang}}, \bibinfo {author} {\bibfnamefont {Y.}~\bibnamefont {Guo}},\ and\
  \bibinfo {author} {\bibfnamefont {D.~L.}\ \bibnamefont {Zhou}},\ }\bibfield
  {title} {\bibinfo {title} {Non-markovian dynamics in a spin star system: the
  failure of thermalisation},\ }\href
  {https://doi.org/10.1140/epjd/e2013-40099-0} {\bibfield  {journal} {\bibinfo
  {journal} {Eur. Phys. J. D}\ }\textbf {\bibinfo {volume} {67}},\ \bibinfo
  {pages} {218} (\bibinfo {year} {2013})}\BibitemShut {NoStop}%
\bibitem [{\citenamefont {Bunyk}\ \emph {et~al.}(2014)\citenamefont {Bunyk},
  \citenamefont {Hoskinson}, \citenamefont {Johnson}, \citenamefont
  {Tolkacheva}, \citenamefont {Altomare}, \citenamefont {Berkley},
  \citenamefont {Harris}, \citenamefont {Hilton}, \citenamefont {Lanting},
  \citenamefont {Przybysz},\ and\ \citenamefont {Whittaker}}]{dwave}%
  \BibitemOpen
  \bibfield  {author} {\bibinfo {author} {\bibfnamefont {P.~I.}\ \bibnamefont
  {Bunyk}}, \bibinfo {author} {\bibfnamefont {E.~M.}\ \bibnamefont
  {Hoskinson}}, \bibinfo {author} {\bibfnamefont {M.~W.}\ \bibnamefont
  {Johnson}}, \bibinfo {author} {\bibfnamefont {E.}~\bibnamefont {Tolkacheva}},
  \bibinfo {author} {\bibfnamefont {F.}~\bibnamefont {Altomare}}, \bibinfo
  {author} {\bibfnamefont {A.~J.}\ \bibnamefont {Berkley}}, \bibinfo {author}
  {\bibfnamefont {R.}~\bibnamefont {Harris}}, \bibinfo {author} {\bibfnamefont
  {J.~P.}\ \bibnamefont {Hilton}}, \bibinfo {author} {\bibfnamefont
  {T.}~\bibnamefont {Lanting}}, \bibinfo {author} {\bibfnamefont {A.~J.}\
  \bibnamefont {Przybysz}},\ and\ \bibinfo {author} {\bibfnamefont
  {J.}~\bibnamefont {Whittaker}},\ }\bibfield  {title} {\bibinfo {title}
  {Architectural considerations in the design of a superconducting quantum
  annealing processor},\ }\href {https://doi.org/10.1109/TASC.2014.2318294}
  {\bibfield  {journal} {\bibinfo  {journal} {IEEE Trans. Appl. Supercond.}\
  }\textbf {\bibinfo {volume} {24}},\ \bibinfo {pages} {1} (\bibinfo {year}
  {2014})}\BibitemShut {NoStop}%
\bibitem [{\citenamefont {Ladd}\ \emph {et~al.}(2010)\citenamefont {Ladd},
  \citenamefont {Jelezko}, \citenamefont {Laflamme}, \citenamefont {Nakamura},
  \citenamefont {Monroe},\ and\ \citenamefont {O'Brien}}]{s-qubit}%
  \BibitemOpen
  \bibfield  {author} {\bibinfo {author} {\bibfnamefont {T.~D.}\ \bibnamefont
  {Ladd}}, \bibinfo {author} {\bibfnamefont {F.}~\bibnamefont {Jelezko}},
  \bibinfo {author} {\bibfnamefont {R.}~\bibnamefont {Laflamme}}, \bibinfo
  {author} {\bibfnamefont {Y.}~\bibnamefont {Nakamura}}, \bibinfo {author}
  {\bibfnamefont {C.}~\bibnamefont {Monroe}},\ and\ \bibinfo {author}
  {\bibfnamefont {J.~L.}\ \bibnamefont {O'Brien}},\ }\bibfield  {title}
  {\bibinfo {title} {Quantum computers},\ }\href
  {https://doi.org/10.1038/nature08812} {\bibfield  {journal} {\bibinfo
  {journal} {Nature}\ }\textbf {\bibinfo {volume} {464}},\ \bibinfo {pages}
  {45} (\bibinfo {year} {2010})}\BibitemShut {NoStop}%
\bibitem [{\citenamefont {Albash}\ and\ \citenamefont
  {Lidar}(2018)}]{coupling}%
  \BibitemOpen
  \bibfield  {author} {\bibinfo {author} {\bibfnamefont {T.}~\bibnamefont
  {Albash}}\ and\ \bibinfo {author} {\bibfnamefont {D.~A.}\ \bibnamefont
  {Lidar}},\ }\bibfield  {title} {\bibinfo {title} {Demonstration of a scaling
  advantage for a quantum annealer over simulated annealing},\ }\href
  {https://doi.org/10.1103/PhysRevX.8.031016} {\bibfield  {journal} {\bibinfo
  {journal} {Phys. Rev. X}\ }\textbf {\bibinfo {volume} {8}},\ \bibinfo {pages}
  {031016} (\bibinfo {year} {2018})}\BibitemShut {NoStop}%
\bibitem [{\citenamefont {Breuer}\ and\ \citenamefont
  {Petruccione}(2002)}]{breuer}%
  \BibitemOpen
  \bibfield  {author} {\bibinfo {author} {\bibfnamefont {H.~P.}\ \bibnamefont
  {Breuer}}\ and\ \bibinfo {author} {\bibfnamefont {F.}~\bibnamefont
  {Petruccione}},\ }\href@noop {} {\emph {\bibinfo {title} {The Theory of Open
  Quantum Systems}}}\ (\bibinfo  {publisher} {Oxford University Press},\
  \bibinfo {address} {Oxford, UK},\ \bibinfo {year} {2002})\BibitemShut
  {NoStop}%
\bibitem [{\citenamefont {Shabani}\ and\ \citenamefont {Neven}(2016)}]{ggl}%
  \BibitemOpen
  \bibfield  {author} {\bibinfo {author} {\bibfnamefont {A.}~\bibnamefont
  {Shabani}}\ and\ \bibinfo {author} {\bibfnamefont {H.}~\bibnamefont
  {Neven}},\ }\bibfield  {title} {\bibinfo {title} {Artificial quantum thermal
  bath: Engineering temperature for a many-body quantum system},\ }\href
  {https://doi.org/10.1103/PhysRevA.94.052301} {\bibfield  {journal} {\bibinfo
  {journal} {Phys. Rev. A}\ }\textbf {\bibinfo {volume} {94}},\ \bibinfo
  {pages} {052301} (\bibinfo {year} {2016})}\BibitemShut {NoStop}%
\bibitem [{\citenamefont {Landsberg}(1956)}]{3rdlaw-2}%
  \BibitemOpen
  \bibfield  {author} {\bibinfo {author} {\bibfnamefont {P.~T.}\ \bibnamefont
  {Landsberg}},\ }\bibfield  {title} {\bibinfo {title} {Foundations of
  thermodynamics},\ }\href {https://doi.org/10.1103/RevModPhys.28.363}
  {\bibfield  {journal} {\bibinfo  {journal} {Rev. Mod. Phys.}\ }\textbf
  {\bibinfo {volume} {28}},\ \bibinfo {pages} {363} (\bibinfo {year}
  {1956})}\BibitemShut {NoStop}%
\bibitem [{\citenamefont {Levy}\ \emph {et~al.}(2012)\citenamefont {Levy},
  \citenamefont {Alicki},\ and\ \citenamefont {Kosloff}}]{3rdlaw-1}%
  \BibitemOpen
  \bibfield  {author} {\bibinfo {author} {\bibfnamefont {A.}~\bibnamefont
  {Levy}}, \bibinfo {author} {\bibfnamefont {R.}~\bibnamefont {Alicki}},\ and\
  \bibinfo {author} {\bibfnamefont {R.}~\bibnamefont {Kosloff}},\ }\bibfield
  {title} {\bibinfo {title} {Quantum refrigerators and the third law of
  thermodynamics},\ }\href {https://doi.org/10.1103/PhysRevE.85.061126}
  {\bibfield  {journal} {\bibinfo  {journal} {Phys. Rev. E}\ }\textbf {\bibinfo
  {volume} {85}},\ \bibinfo {pages} {061126} (\bibinfo {year}
  {2012})}\BibitemShut {NoStop}%
\bibitem [{\citenamefont {Kosloff}\ and\ \citenamefont
  {Rezek}(2017)}]{ottorev}%
  \BibitemOpen
  \bibfield  {author} {\bibinfo {author} {\bibfnamefont {R.}~\bibnamefont
  {Kosloff}}\ and\ \bibinfo {author} {\bibfnamefont {Y.}~\bibnamefont
  {Rezek}},\ }\bibfield  {title} {\bibinfo {title} {The quantum harmonic otto
  cycle},\ }\href {https://doi.org/10.3390/e19040136} {\bibfield  {journal}
  {\bibinfo  {journal} {Entropy}\ }\textbf {\bibinfo {volume} {19(4)}},\
  \bibinfo {pages} {136} (\bibinfo {year} {2017})}\BibitemShut {NoStop}%
\bibitem [{\citenamefont {Roßnagel}\ \emph {et~al.}(2016)\citenamefont
  {Roßnagel}, \citenamefont {Dawkins}, \citenamefont {Tolazzi}, \citenamefont
  {Abah}, \citenamefont {Lutz}, \citenamefont {Schmidt-Kaler},\ and\
  \citenamefont {Singer}}]{iontrap}%
  \BibitemOpen
  \bibfield  {author} {\bibinfo {author} {\bibfnamefont {J.}~\bibnamefont
  {Roßnagel}}, \bibinfo {author} {\bibfnamefont {S.}~\bibnamefont {Dawkins}},
  \bibinfo {author} {\bibfnamefont {K.}~\bibnamefont {Tolazzi}}, \bibinfo
  {author} {\bibfnamefont {O.}~\bibnamefont {Abah}}, \bibinfo {author}
  {\bibfnamefont {E.}~\bibnamefont {Lutz}}, \bibinfo {author} {\bibfnamefont
  {F.}~\bibnamefont {Schmidt-Kaler}},\ and\ \bibinfo {author} {\bibfnamefont
  {K.~A.}\ \bibnamefont {Singer}},\ }\bibfield  {title} {\bibinfo {title} {A
  single-atom heat engine},\ }\href {https://doi.org/10.1126/science.aad6320}
  {\bibfield  {journal} {\bibinfo  {journal} {Science}\ }\textbf {\bibinfo
  {volume} {352}},\ \bibinfo {pages} {325} (\bibinfo {year}
  {2016})}\BibitemShut {NoStop}%
\bibitem [{\citenamefont {Abah}\ \emph {et~al.}(2012)\citenamefont {Abah},
  \citenamefont {Roßnagel}, \citenamefont {Jacob}, \citenamefont {Deffner},
  \citenamefont {Schmidt-Kaler}, \citenamefont {Singer},\ and\ \citenamefont
  {Lutz}}]{nonad-otto1}%
  \BibitemOpen
  \bibfield  {author} {\bibinfo {author} {\bibfnamefont {O.}~\bibnamefont
  {Abah}}, \bibinfo {author} {\bibfnamefont {J.}~\bibnamefont {Roßnagel}},
  \bibinfo {author} {\bibfnamefont {G.}~\bibnamefont {Jacob}}, \bibinfo
  {author} {\bibfnamefont {S.}~\bibnamefont {Deffner}}, \bibinfo {author}
  {\bibfnamefont {F.}~\bibnamefont {Schmidt-Kaler}}, \bibinfo {author}
  {\bibfnamefont {K.}~\bibnamefont {Singer}},\ and\ \bibinfo {author}
  {\bibfnamefont {E.}~\bibnamefont {Lutz}},\ }\bibfield  {title} {\bibinfo
  {title} {Single-ion heat engine at maximum power},\ }\href
  {https://doi.org/10.1103/PhysRevLett.109.203006} {\bibfield  {journal}
  {\bibinfo  {journal} {Phys. Rev. Lett}\ }\textbf {\bibinfo {volume} {109}},\
  \bibinfo {pages} {203006} (\bibinfo {year} {2012})}\BibitemShut {NoStop}%
\bibitem [{\citenamefont {Roßnagel}\ \emph {et~al.}(2014)\citenamefont
  {Roßnagel}, \citenamefont {Abah}, \citenamefont {Schmidt-Kaler},
  \citenamefont {Singer},\ and\ \citenamefont {Lutz}}]{ottosq1}%
  \BibitemOpen
  \bibfield  {author} {\bibinfo {author} {\bibfnamefont {J.}~\bibnamefont
  {Roßnagel}}, \bibinfo {author} {\bibfnamefont {O.}~\bibnamefont {Abah}},
  \bibinfo {author} {\bibfnamefont {F.}~\bibnamefont {Schmidt-Kaler}}, \bibinfo
  {author} {\bibfnamefont {K.}~\bibnamefont {Singer}},\ and\ \bibinfo {author}
  {\bibfnamefont {E.}~\bibnamefont {Lutz}},\ }\bibfield  {title} {\bibinfo
  {title} {Nanoscale heat engine beyond the carnot limit},\ }\href
  {https://doi.org/10.1103/PhysRevLett.112.030602} {\bibfield  {journal}
  {\bibinfo  {journal} {Phys. Rev. Lett.}\ }\textbf {\bibinfo {volume} {112}},\
  \bibinfo {pages} {030602} (\bibinfo {year} {2014})}\BibitemShut {NoStop}%
\bibitem [{\citenamefont {Klaers}\ \emph {et~al.}(2017)\citenamefont {Klaers},
  \citenamefont {Faelt}, \citenamefont {Imamo{\u g}lu},\ and\ \citenamefont
  {Togan}}]{ottosq2}%
  \BibitemOpen
  \bibfield  {author} {\bibinfo {author} {\bibfnamefont {J.}~\bibnamefont
  {Klaers}}, \bibinfo {author} {\bibfnamefont {S.}~\bibnamefont {Faelt}},
  \bibinfo {author} {\bibfnamefont {A.}~\bibnamefont {Imamo{\u g}lu}},\ and\
  \bibinfo {author} {\bibfnamefont {E.}~\bibnamefont {Togan}},\ }\bibfield
  {title} {\bibinfo {title} {Squeezed thermal reservoirs as a resource for a
  nanomechanical engine beyond the carnot limit},\ }\href
  {https://doi.org/10.1103/PhysRevX.7.031044} {\bibfield  {journal} {\bibinfo
  {journal} {Phys. Rev. X}\ }\textbf {\bibinfo {volume} {7}},\ \bibinfo {pages}
  {031044} (\bibinfo {year} {2017})}\BibitemShut {NoStop}%
\bibitem [{\citenamefont {Manzano}\ \emph {et~al.}(2016)\citenamefont
  {Manzano}, \citenamefont {Galve}, \citenamefont {Zambrini},\ and\
  \citenamefont {Parrondo}}]{ottosq3}%
  \BibitemOpen
  \bibfield  {author} {\bibinfo {author} {\bibfnamefont {G.}~\bibnamefont
  {Manzano}}, \bibinfo {author} {\bibfnamefont {F.}~\bibnamefont {Galve}},
  \bibinfo {author} {\bibfnamefont {R.}~\bibnamefont {Zambrini}},\ and\
  \bibinfo {author} {\bibfnamefont {J.~M.}\ \bibnamefont {Parrondo}},\
  }\bibfield  {title} {\bibinfo {title} {Entropy production and thermodynamic
  power of the squeezed thermal reservoir},\ }\href
  {https://doi.org/10.1103/PhysRevE.93.052120} {\bibfield  {journal} {\bibinfo
  {journal} {Phys. Rev. E}\ }\textbf {\bibinfo {volume} {93}},\ \bibinfo
  {pages} {052120} (\bibinfo {year} {2016})}\BibitemShut {NoStop}%
\bibitem [{\citenamefont {Metcalf}\ \emph {et~al.}(2020)\citenamefont
  {Metcalf}, \citenamefont {Moussa}, \citenamefont {de~Jong},\ and\
  \citenamefont {Sarovar}}]{metcalf2020}%
  \BibitemOpen
  \bibfield  {author} {\bibinfo {author} {\bibfnamefont {M.}~\bibnamefont
  {Metcalf}}, \bibinfo {author} {\bibfnamefont {J.~E.}\ \bibnamefont {Moussa}},
  \bibinfo {author} {\bibfnamefont {W.~A.}\ \bibnamefont {de~Jong}},\ and\
  \bibinfo {author} {\bibfnamefont {M.}~\bibnamefont {Sarovar}},\ }\bibfield
  {title} {\bibinfo {title} {Engineered thermalization and cooling of quantum
  many-body systems},\ }\href
  {https://doi.org/10.1103/PhysRevResearch.2.023214} {\bibfield  {journal}
  {\bibinfo  {journal} {Phys. Rev. Research}\ }\textbf {\bibinfo {volume}
  {2}},\ \bibinfo {pages} {023214} (\bibinfo {year} {2020})}\BibitemShut
  {NoStop}%
\bibitem [{\citenamefont {Jordan}\ and\ \citenamefont
  {B\"{u}ttiker}(2004)}]{entq}%
  \BibitemOpen
  \bibfield  {author} {\bibinfo {author} {\bibfnamefont {A.~N.}\ \bibnamefont
  {Jordan}}\ and\ \bibinfo {author} {\bibfnamefont {M.}~\bibnamefont
  {B\"{u}ttiker}},\ }\bibfield  {title} {\bibinfo {title} {Entanglement
  energetics at zero temperature},\ }\href
  {https://doi.org/10.1103/PhysRevLett.92.247901} {\bibfield  {journal}
  {\bibinfo  {journal} {Phys. Rev. Lett.}\ }\textbf {\bibinfo {volume} {92}},\
  \bibinfo {pages} {247901} (\bibinfo {year} {2004})}\BibitemShut {NoStop}%
\bibitem [{\citenamefont {Ingold}\ \emph {et~al.}(2009)\citenamefont {Ingold},
  \citenamefont {H{\"a}nggi},\ and\ \citenamefont {Talkner}}]{sha1}%
  \BibitemOpen
  \bibfield  {author} {\bibinfo {author} {\bibfnamefont {G.-L.}\ \bibnamefont
  {Ingold}}, \bibinfo {author} {\bibfnamefont {P.}~\bibnamefont {H{\"a}nggi}},\
  and\ \bibinfo {author} {\bibfnamefont {P.}~\bibnamefont {Talkner}},\
  }\bibfield  {title} {\bibinfo {title} {Specific heat anomalies of open
  quantum systems},\ }\href {https://doi.org/10.1103/PhysRevE.79.061105}
  {\bibfield  {journal} {\bibinfo  {journal} {Phys. Rev. E}\ }\textbf {\bibinfo
  {volume} {79}},\ \bibinfo {pages} {061105} (\bibinfo {year}
  {2009})}\BibitemShut {NoStop}%
\bibitem [{\citenamefont {Hasegawa}(2011)}]{sha2}%
  \BibitemOpen
  \bibfield  {author} {\bibinfo {author} {\bibfnamefont {H.}~\bibnamefont
  {Hasegawa}},\ }\bibfield  {title} {\bibinfo {title} {Specific heat anomalies
  of small quantum systems subjected to finite baths},\ }\href
  {https://doi.org/10.1063/1.3669485} {\bibfield  {journal} {\bibinfo
  {journal} {J. Math. Phys.}\ }\textbf {\bibinfo {volume} {52}},\ \bibinfo
  {pages} {123301} (\bibinfo {year} {2011})}\BibitemShut {NoStop}%
\end{thebibliography}%
\end{document}